\documentclass[useAMX,usenatbib,usegraphicx]{mn2e}

\usepackage{times}

\newcommand{\JHK}{{JHK_\mathrm{s}}}
\newcommand{\Ks}{{K_\mathrm{s}}}
\newcommand{\muLMC}{{\mu_\mathrm{LMC}}}

\title[P-L Relations for Type II Cepheids]{Period-Luminosity Relations for Type II Cepheids
and their Application}
\author[N. Matsunaga, M.~W. Feast and J.~W. Menzies]
{Noriyuki Matsunaga$^{1}$\thanks{An e-mail address and the current address of NM are as follows: matsunaga@ioa.s.u-tokyo.ac.jp, Institute of Astronomy, University of Tokyo, 2-21-1 Osawa, Mitaka, Tokyo 181-0015, Japan},
Michael W. Feast$^{2,3}$, and John W. Menzies$^{3}$\\
$^{1}$ Department of Astronomy, Kyoto University, Kitashirakawa-Oiwake-cho,
Sakyo-ku, Kyoto 606-8502, Japan;\\
Research Fellow of the Japan Society for the Promotion of Science\\
$^{2}$ Department of Astronomy, University of Cape Town, 
7701, Rondebosch, South Africa\\
$^{3}$ South African Astronomical Observatory, PO Box 9, 7935, Observatory,
South Africa\\}

\begin{document}

\date{Accepted 29 April 2009. Received 29 April 2009; in original form 24 March 2009}

\pagerange{\pageref{firstpage}--\pageref{lastpage}} \pubyear{2009}

\maketitle

\label{firstpage}

\begin{abstract}
${\JHK}$ magnitudes corrected to mean intensity are estimated for
LMC type II Cepheids in the OGLE-III survey. Period-luminosity (PL)
relations are derived in $\JHK$ as well as in a reddening-free $VI$
parameter. Within the uncertainties the BL Her stars ($P < 4$~d)
and the W Vir stars ($P = 4$ to 20~d) are co-linear in these PL relations.
The slopes of the infrared relations agree with those found previously for
type II Cepheids in globular clusters within the uncertainties.
Using the pulsation parallaxes of V553~Cen and SW~Tau the data lead to
an LMC modulus uncorrected for any metallicity effects of $18.46\pm 0.10$~mag.
The type II Cepheids in the second-parameter globular cluster, NGC~6441,
show a PL($VI$) relation of the same slope as that in the LMC and this
leads to a cluster distance modulus of $15.46\pm 0.11$~mag, confirming
the hypothesis that the RR Lyrae variables in this cluster are overluminous
for their metallicity. It is suggested that the Galactic variable
$\kappa$ Pav is a member of the peculiar W Vir class found by
the OGLE-III group in the LMC. Low-resolution spectra of OGLE-III type II
Cepheids with $P > 20$~d (RV Tau stars) show that a high proportion
have TiO bands; only one has been found showing $\rm C_{2}$. The LMC
RV Tau stars, as a group, are not co-linear with the shorter-period type II
Cepheids in the infrared PL relations in marked contrast to such stars in
globular clusters. Other differences between LMC, globular cluster
and Galactic field type II Cepheids are noted in period distribution
and infrared colours.
\end{abstract}

\section{Introduction}
\label{sec:Intro}
   Type II Cepheids (CephIIs) have periods in the same range as classical 
Cepheids but are lower-mass stars
belonging to disc and halo populations. They are conventionally divided 
into
three period groups; BL Her stars (BL) at short periods, W Vir stars (WV) at
intermediate periods, and RV Tau stars (RV) at the longest periods. The period
divisions tend to be somewhat arbitrary. In a recent paper to which
considerable reference will be made,
\citeauthor{Soszynski-2008} (\citeyear{Soszynski-2008}, hereafter S08) 
adopt divisions at 4 and 20~d,
and we follow these thresholds here.
The RV stars tend to show alternating
deep and shallow minima (and this is often taken as a defining characteristic),
but the single period of the RVs will be used in this paper as in S08.
In addition to the BL, WV, and RV stars, S08 define an additional
class of peculiar W Vir (pW) stars which will be discussed below.

Most workers have accepted the evolutionary scheme elaborated by 
Gingold (\citeyear{Gingold-1976}, \citeyear{Gingold-1985}).
In this the BL stars are evolving from the (blue) horizontal branch
towards the lower asymptotic giant branch (AGB).
The WV stars are on loops to the blue from the AGB
and the RV stars are moving to the blue in a post-AGB phase. 
\citeauthor{Matsunaga-2006} (\citeyear{Matsunaga-2006}, hereafter M06) 
showed that the CephIIs in globular clusters defined narrow period-luminosity
(PL) relations in the near-infrared bands, $\JHK$, with little evidence
for a metallicity
effect in these relations. This suggests that these stars may be useful 
distance indicators for disc and halo populations. Pulsation parallaxes of 
Galactic
CephIIs were used by 
\citeauthor{Feast-2008} (\citeyear{Feast-2008}, hereafter F08) 
to calibrate these  cluster PL relations
and to discuss the distances of the LMC and the Galactic centre. 

The present 
paper discusses the CephIIs in the LMC based on the recent optical OGLE-III
survey (S08) and the results of the 
near-infrared survey with
the Infrared Survey Facility (IRSF).
The 
CephIIs in the second-parameter cluster NGC~6441 and those around
the Galactic centre are also discussed as well
as the nature of the CephII variable $\kappa$~Pav.   

\section{Infrared Photometry}
\label{sec:Photometry}
S08 catalogued 197 CephII variables in the LMC.
We searched for near-infrared counterparts in the IRSF catalogue
\citep{Kato-2007}. This is a point-source catalogue  in $\JHK$ of
$\rm 40~deg^{2}$ of the LMC, $\rm 11~deg^{2}$ of the SMC, and
$\rm 4~deg^{2}$ of the Magellanic Bridge. The catalogue is based
on simultaneous images in $\JHK$ obtained with the 1.4-m
IRSF telescope and
the Simultaneous 3-colour Infrared Imager for Unbiased Survey (SIRIUS)
based at the South African Astronomical Observatory (SAAO), Sutherland.
The 10~$\sigma$ limiting magnitudes of the survey are 18.8, 17.8 and
16.6~mag $J$, $H$, and $\Ks$.
The catalogue reaches fainter and has higher resolution than
the
point-source catalogue of the 2~Micron All Sky Survey
(2MASS,
\citealt{Skrutskie-2006}) in the region of the Magellanic Clouds.

We found matches for 188 S08 CephII sources
with a tolerance of 0.5~arcsec. 
The differences in coordinate are small between the catalogues
with a standard deviation of less than 0.1~arcsec 
in both Right Ascension and Declination.
Among nine sources without IRSF counterparts, seven (1--3, 194--197)
are located outside of the IRSF survey field.
There are no counterparts around the two BL stars (69 and 123)
probably because they are too faint. These variables
actually have the shortest periods among our sample.
On the other hand, we found two IRSF counterparts
for 15 S08 sources:
these IRSF counterparts were detected in neighbouring fields
of the IRSF survey as listed in Table \ref{tab:double}.

IRSF magnitudes in all three bands are
not available for nine BLs and one pW.
In most of the cases, $\Ks$-band magnitudes are missing because
of the faintness of the short-period BLs.
$H$-band magnitudes are unavailable for No.\  40 (pW)
and 144 (BL); the reason for this is unclear.

Since S08 give the periods and dates of maximum light in $I$ (= phase zero),
we can derive the phases (between zero and one) of the single
$\JHK$ observations. Using the $I$-band light 
curves we obtain the value of $I$ at that phase and its difference from the
intensity mean value, $\langle I\rangle$.
Assuming that the light curves in $\JHK$ are similar to those in $I$, we 
obtain an estimate of the mean (phase-corrected) magnitude in each infrared 
band.

\begin{figure*}
\begin{center}
\begin{minipage}{170mm}
\begin{center}
\includegraphics[clip,width=0.95\hsize]{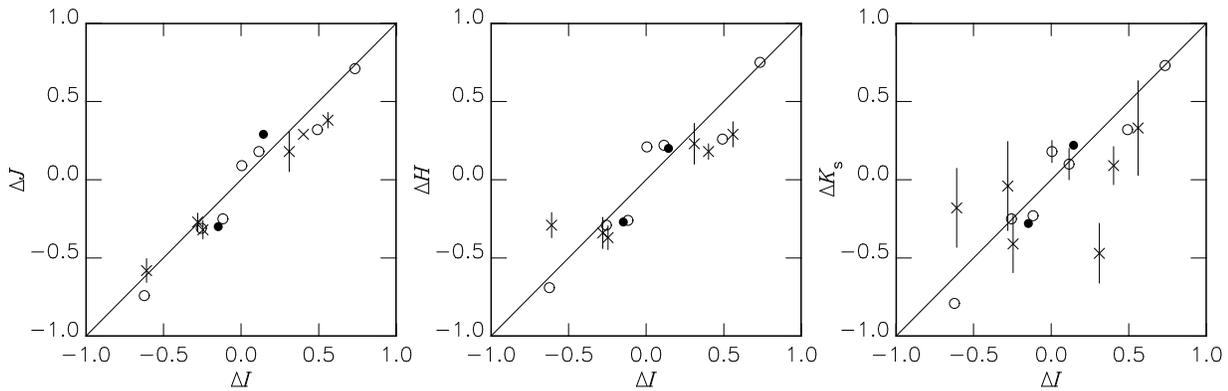}
\end{center}
\caption{
Differences of $\JHK$ magnitudes at two epochs of the IRSF observations
are plotted against those of $I$-band magnitudes estimated from
the OGLE light curves. Crosses indicate BLs, open circles WVs,
and filled circles RVs. The error bars are omitted when their
sizes are smaller than the symbols.
\label{fig:compAmp}}
\end{minipage}
\end{center}
\end{figure*}

In order to check the validity of the above assumption,
we use the S08 sources with two IRSF counterparts.
There are 15 such sources (2 RV, 7 WV, 6 BL).
We list the dates, phases, and observed magnitudes
from the two IRSF measurements in Table \ref{tab:double}.
For each phase at which the IRSF survey was conducted,
we also estimate a predicted $I$-band magnitude
by taking a mean of the $I$-band measurements which were made
within $\pm 0.05$ of the phase of the IRSF survey.
Variations between two IRSF measurements
($\Delta J$, $\Delta H$, $\Delta \Ks$)
have the same signs as the difference between the predicted $I$-band
magnitudes ($\Delta I$)
for each object. Fig.~\ref{fig:compAmp} plots the variations
of the IRSF measurements against $\Delta I$.
This clearly shows the variations in $\JHK$ reasonably agree with
those in $I$,
except for the $\Delta \Ks$ values for BLs (crosses) which have
large error bars. That is, these observations are consistent
with the assumption that the $\JHK$ light curves are, to a first
approximation, similar
to those at $I$ and have about the same amplitude.

\begin{table*}
\begin{minipage}{170mm}
\caption{
The LMC CephIIs with two IRSF measurements. Modified Julian Dates (MJD),
pulsation phase of the observations, and $\JHK$ magnitudes
and their errors
are listed for each IRSF entry. Also indicated are $I$-band
magnitudes at the phases of the IRSF measurements
based on the OGLE light curves.
\label{tab:double}}
\begin{center}
\begin{tabular}{ccccccccccccc}
\hline
OGLE- & Type & $\log P$ & \multicolumn{9}{c}{IRSF counterpart} & Expected $I$ \\
ID  &  &  & IRSF-Field & MJD(obs) & Phase & $J$ &$E_J$ & $H$ &$E_H$ & $\Ks$ &$E_K$ & \\
\hline
10 & BLHer & 0.17695 & LMC0456-6840I & 53062.869 & 0.681 & 17.71 & 0.04 & 17.35 & 0.04 & 17.23 & 0.25 & 18.26 \\
10 & & & LMC0452-6840G & 53049.895 & 0.049 & 17.33 & 0.03 & 17.06 & 0.04 & 16.90 & 0.17 & 17.70 \\
11 & RVTau & 1.59391 & LMC0453-6740G & 52683.765 & 0.578 & 13.71 & 0.03 & 13.32 & 0.03 & 13.25 & 0.02 & 14.17 \\
11 & & & LMC0454-6720A & 53017.960 & 0.091 & 13.42 & 0.01 & 13.12 & 0.01 & 13.03 & 0.02 & 14.03 \\
22 & WVir & 1.03006 & LMC0458-7040G & 53497.695 & 0.076 & 15.44 & 0.02 & 15.05 & 0.02 & 15.04 & 0.03 & 16.13 \\
22 & & & LMC0502-7040I & 53117.759 & 0.623 & 15.75 & 0.02 & 15.34 & 0.02 & 15.29 & 0.03 & 16.38 \\
38 & WVir & 0.60354 & LMC0503-6840A & 53123.751 & 0.964 & 15.94 & 0.02 & 15.87 & 0.02 & 15.84 & 0.05 & 16.10 \\
38 & & & LMC0507-6840C & 53341.916 & 0.320 & 15.85 & 0.01 & 15.66 & 0.02 & 15.66 & 0.05 & 16.09 \\
50 & RVTau & 1.54093 & LMC0511-6840C & 53051.855 & 0.304 & 14.35 & 0.02 & 14.03 & 0.02 & 13.82 & 0.02 & 14.93 \\
50 & & & LMC0511-6900I & 52612.916 & 0.672 & 14.65 & 0.03 & 14.30 & 0.02 & 14.10 & 0.02 & 15.07 \\
59 & WVir & 1.22365 & LMC0510-7040G & 53322.877 & 0.012 & 14.68 & 0.01 & 14.36 & 0.01 & 14.22 & 0.02 & 15.28 \\
59 & & & LMC0514-7040I & 52714.762 & 0.677 & 15.42 & 0.02 & 15.05 & 0.02 & 15.01 & 0.04 & 15.90 \\
76 & BLHer & 0.32311 & LMC0518-6820I & 52657.914 & 0.487 & 17.19 & 0.03 & 16.77 & 0.03 & 16.64 & 0.09 & 17.85 \\
76 & & & LMC0514-6820G & 52660.786 & 0.852 & 17.51 & 0.05 & 17.14 & 0.07 & 17.05 & 0.16 & 18.09 \\
100 & WVir & 0.87105 & LMC0522-7020F & 53355.891 & 0.554 & 16.31 & 0.02 & 16.01 & 0.02 & 16.01 & 0.08 & 16.83 \\
100 & & & LMC0522-7020E & 52700.891 & 0.411 & 16.13 & 0.02 & 15.79 & 0.02 & 15.91 & 0.06 & 16.71 \\
105 & BLHer & 0.17298 & LMC0522-7020E & 52700.891 & 0.040 & 17.05 & 0.03 & 16.84 & 0.03 & 16.74 & 0.14 & 17.32 \\
105 & & & LMC0522-7020H & 52701.758 & 0.622 & 17.63 & 0.07 & 17.13 & 0.07 & 16.92 & 0.21 & 17.93 \\
118 & WVir & 1.10376 & LMC0525-6800B & 52673.837 & 0.240 & 15.44 & 0.02 & 15.00 & 0.02 & 14.90 & 0.03 & 16.30 \\
118 & & & LMC0525-6820H & 52683.781 & 0.023 & 15.12 & 0.03 & 14.74 & 0.02 & 14.58 & 0.03 & 15.81 \\
122 & BLHer & 0.18715 & LMC0525-6840H & 52683.954 & 0.007 & 17.46 & 0.04 & 17.14 & 0.04 & 17.10 & 0.22 & 18.05 \\
122 & & & LMC0525-6820B & 52675.910 & 0.779 & 17.73 & 0.04 & 17.48 & 0.08 & 17.14 & 0.18 & 18.33 \\
138 & BLHer & 0.14414 & LMC0529-6840A & 52684.777 & 0.524 & 17.29 & 0.10 & 17.43 & 0.10 & 16.52 & 0.13 & 18.23 \\
138 & & & LMC0529-6900G & 52364.819 & 0.931 & 17.11 & 0.08 & 17.20 & 0.07 & 16.99 & 0.14 & 17.92 \\
143 & WVir & 1.16347 & LMC0529-6920G & 52363.769 & 0.569 & 15.57 & 0.02 & 15.27 & 0.02 & 15.17 & 0.04 & 16.21 \\
143 & & & LMC0533-6920I & 53331.899 & 0.014 & 14.86 & 0.03 & 14.52 & 0.02 & 14.44 & 0.02 & 15.47 \\
146 & WVir & 1.00344 & LMC0533-6840C & 52687.758 & 0.228 & 15.63 & 0.02 & 15.25 & 0.02 & 15.12 & 0.03 & 16.39 \\
146 & & & LMC0533-6900I & 52286.935 & 0.463 & 15.88 & 0.01 & 15.51 & 0.02 & 15.35 & 0.03 & 16.50 \\
148 & BLHer & 0.42679 & LMC0533-6920C & 52969.042 & 0.572 & 17.13 & 0.02 & 16.85 & 0.02 & 16.66 & 0.07 & 17.73 \\
148 & & & LMC0533-6940I & 52226.973 & 0.824 & 16.84 & 0.02 & 16.67 & 0.04 & 16.57 & 0.10 & 17.33 \\
\hline
\end{tabular}
\end{center}
\end{minipage}
\end{table*}

Light curves of RV stars often have a large scatter,
and it is difficult to make satisfactory phase corrections.
We selected eight RVs which have good light curves,
i.e. 5, 58, 104, 115, 125, 135, 169, and 192,
and use these light curves to make phase corrections.
For pW, we do not try to correct the phase effect,
and they are not included in the following discussions of the PL relation.

We here present a catalogue of the S08 sources with the IRSF counterparts
in Table \ref{tab:catalogue}, where
the $\JHK$ values are
those listed in the IRSF catalogue \citep{Kato-2007}.
The indicated errors, $E_J, E_H$, and $E_K$, are taken from
their catalogue.
The quantity $\delta_{\phi}$ is the correction which must be added
to the $I$ magnitude at the epoch of the IRSF observation to correct it to
the intensity mean magnitude and is the value we
adopt to correct the infrared magnitudes to means.
Note that we list
those with two IRSF counterparts in neighbouring fields twice.

\begin{table*}
\begin{minipage}{170mm}
\caption{
The first ten lines of the catalogue of S08 sources with IRSF counterparts.
This is a sample of the full version, which will be available in the
online version of this journal.
Modified Julian Dates (MJD),
pulsation phase of the observations, $\JHK$ magnitudes, and their errors
are listed for each IRSF measurement as well as the OGLE-IDs,
types, and periods.
Shifts for the phase corrections obtained from
the $I$-band light curves are also listed if available.
Nine S08 sources are absent because their IRSF
counterparts were not found, and 15 S08 sources are listed twice
because they are identified with two counterparts from
neighbouring fields of the IRSF survey.
\label{tab:catalogue}}
\begin{center}
\begin{tabular}{ccccccccccccc}
\hline
 OGLE- & Type & $\log P$ & \multicolumn{9}{c}{IRSF counterpart} & $\delta _{\phi}$ \\
 ID & & & IRSF-Field & MJD(obs) & Phase & $J$ & $E_J$ & $H$ & $E_H$ & $\Ks$ & $E_K$ & \\
\hline
4   & BLHer & 0.28240 & LMC0446-6800C & 52645.086 & 0.121 & 16.64 & 0.04 & 16.33 & 0.06 & 16.27 & 0.11 & $ 0.203$ \\
5   & RVTau & 1.52095 & LMC0447-7000G & 53037.995 & 0.161 & 13.82 & 0.02 & 13.41 & 0.02 & 13.28 & 0.03 & $ 0.201$ \\
6   & BLHer & 0.03660 & LMC0450-6720B & 53031.900 & 0.273 & 17.63 & 0.04 & 17.43 & 0.07 & 17.32 & 0.22 & $ 0.047$ \\
7   & BLHer & 0.09435 & LMC0452-6920F & 53044.963 & 0.224 & 17.50 & 0.04 & 17.25 & 0.06 &    -- &   -- & $-0.005$ \\
8   & BLHer & 0.24207 & LMC0451-7000E & 53308.910 & 0.293 & 17.25 & 0.02 & 17.34 & 0.06 & 17.31 & 0.17 & $ 0.130$ \\
9   & BLHer & 0.24584 & LMC0454-6700C & 53365.791 & 0.222 & 17.21 & 0.04 & 16.82 & 0.06 & 16.92 & 0.19 & $ 0.069$ \\
10  & BLHer & 0.17695 & LMC0456-6840I & 53062.869 & 0.681 & 17.71 & 0.04 & 17.35 & 0.07 & 17.23 & 0.25 & $-0.281$ \\
10  & BLHer & 0.17695 & LMC0452-6840G & 53049.895 & 0.049 & 17.33 & 0.03 & 17.06 & 0.04 & 16.90 & 0.17 & $ 0.278$ \\
11  & RVTau & 1.59391 & LMC0453-6740G & 52683.765 & 0.578 & 13.71 & 0.03 & 13.32 & 0.02 & 13.25 & 0.02 & -- \\
11  & RVTau & 1.59391 & LMC0454-6720A & 53017.960 & 0.091 & 13.42 & 0.01 & 13.12 & 0.01 & 13.03 & 0.02 & -- \\
\hline
\end{tabular}
\end{center}
\end{minipage}
\end{table*}

\section{Period-luminosity relations}
\label{sec:PLR}

\subsection{General and Optical Relations\label{sec:optPLR}}
   In this section we discuss  the optical PL relations using the data in 
S08. Based on notes given by S08 the following stars were omitted in all
solutions, optical and infrared; 88, 153, 166, 185 (blends); 21, 23, 52,
77, 84, 93, 98 (eclipsing), 50 (too blue), 108 (low amplitude), 113 (scatter
in light curve), 51 (variable amplitude), and 150 (variable mean magnitude).   

The general features of CephII optical period-luminosity diagrams are
well shown in fig.~1 of S08. The plots of $V$ and $I$ against $\log P$ show 
considerable scatter and clear non-linearity.
Introduction of a colour term $[W = I-R(V-I)]$
with $R = 1.55$ produces a rather narrow and nearly linear PL($W$) relation
for the BL and WV stars although there is still scatter amongst the 
RVs. 

S08 recognize a class of peculiar W Vir (pW) stars. These have distinctive
light curves and a high proportion are binaries. Many of these lie above
(brighter than) the normal WV stars in the various PL plots. In the solutions below all the pW stars 
are omitted. There are also a few stars in the
BL period range which in the PL($W$) plot lie brighter than the bulk of the BL
stars and in the region occupied by anomalous Cepheids. We have chosen to
omit these stars (107, 114, 142, 153, 166) in our work. 

If the value of $R$ is correctly chosen, $W$ is a reddening-free parameter.
\citet{Udalski-1999} found $R=1.55$ as appropriate to the OGLE-II
photometry from the results of
Schlegel, Finkbeiner, \& Davis (1998). 
On the other hand, a value of $R=1.45$ has frequently been used
(e.g.\  \citealt{Freedman-2001}; \citealt{vanLeeuwen-2007}).
This latter value is based on the extinction law of
Cardelli, Clayton, \& Mathis (1989). 
We give below results based on both values of $R$. In work with very heavily
reddened stars the exact value of $R$ may become important.

For 55  BL stars we find:
\begin{eqnarray}
W_{1} = I - 1.55 (V-I) = -2.598(\pm 0.094)(\log P -0.3) \nonumber\\ 
+ 16.597(\pm 0.017),
 (\sigma = 0.104), \label{eq:W1}
\end{eqnarray}
and
\begin{eqnarray}
W_{2} = I - 1.45(V-I) = -2.572(\pm 0.093)(\log P - 0.3) \nonumber\\
+ 16.665(\pm 0.016),
(\sigma  = 0.103). \label{eq:W2}
\end{eqnarray}
For 76 WV stars we find:
\begin{eqnarray}
W_{3} = I - 1.55(V-I) = -2.564(\pm 0.073)(\log P - 1.2) \nonumber\\
+ 14.333(\pm 0.019),
(\sigma = 0.108), \label{eq:W3}
\end{eqnarray}
and
\begin{eqnarray}
W_{4} = I - 1.45(V-I) = -2.551(\pm 0.073)(\log P - 1.2) \nonumber\\
+ 14.431(\pm 0.019),
(\sigma = 0.105). \label{eq:W4}
\end{eqnarray}
The slopes and (effective) zero-points do not differ significantly
between the solutions for BL and WV,
so that we give solutions for BL+WV stars:
\begin{eqnarray}
W_{5} = I - 1.55(V-I) = -2.521(\pm 0.022)(\log P - 1.2) \nonumber\\
+ 14.339(\pm 0.015),
(\sigma = 0.105), \label{eq:W5}
\end{eqnarray}
and
\begin{eqnarray}
W_{6} = I - 1.45(V-I) = -2.486(\pm 0.022)(\log P - 1.2) \nonumber\\
+ 14.440(\pm 0.15),
(\sigma = 0.106). \label{eq:W6}
\end{eqnarray}
These relations are narrow ($\sigma \sim 0.1$) as expected from fig.~1
of S08.

Use of the PL($W$) relation brings stars that lie below the 
bulk of the CephIIs in the $\log P$-$V$~(and $I$) diagrams  
and are redder than the bulk of the CephIIs PL
into agreement with the others as expected for reddening-free relations.
In addition we can expect that, as with classical Cepheids, CephIIs occupy
an instability strip of finite width in temperature (i.e.\  colour). Thus
a PL relation has a finite scatter and a colour term 
is needed to reduce this. In the case of the classical Cepheids the 
coefficient of the required colour term is about 1.45 
(\citealt{Udalski-1999}, \citealt{vanLeeuwen-2007}) so that $W$ corrects 
in that
case for both reddening and strip width. 

As a first approximation we might
expect the same value of $R$ 
to apply in the case of the CephIIs. One can test this 
by looking at the residual from the PL($W$) relation of the bluest stars,
since it seems unreasonable to assume that the bulk of the 
BL stars which scatter around $(V-I)\sim 0.7$,
are heavily reddened by interstellar extinction. There
are eight BL bluer than $(V-I) = 0.5$ which were not discarded for any of the
reasons given above (stars 6, 20, 41, 71, 89, 102, 136, and 145).
Their residuals from the above equation for $W_{2}$ are
$0.000$, $+0.084$,
$-0.029$, $-0.145$, $+0.100$, $-0.081$, $+0.305$, and $-0.066$,
respectively,
giving a mean of $+0.021\pm 0.050$ or $-0.020\pm 0.033$ if No.\  136 is omitted.
The residual for 136 is rather large and this is the bluest star in the
sample. 
It seems possible that this might be a blend or a binary like some
of the rejected stars, but this is not certain. 

Taking the above results together with the 
colour-magnitude diagram of S08 (their fig.~2) strongly suggests that the
BL stars occupy an instability strip with 
a width of $\sim 0.4$~mag in $V-I$.
The exact limits to the strip are not clearly
defined. However it is important for the zero-point calibration discussed in
section \ref{sec:Application} that for an intrinsic width of
this order the PL($W$) relation
remains narrow, indicating that in this case (as for classical Cepheid)
the relation corrects well for both reddening and intrinsic colour variation.

Fig.~2 of S08 also
suggests that, if the pW stars are omitted, most of the WVs occupy a narrower
instability strip than the BLs, though it is uncertain whether or not
the colours of some of the redder WVs are due to interstellar reddening or are
intrinsic.

\subsection{Infrared PL relations\label{sec:irPLR}}

\begin{figure*}
\begin{minipage}{140mm}
\begin{center}
\includegraphics[clip,width=0.85\hsize]{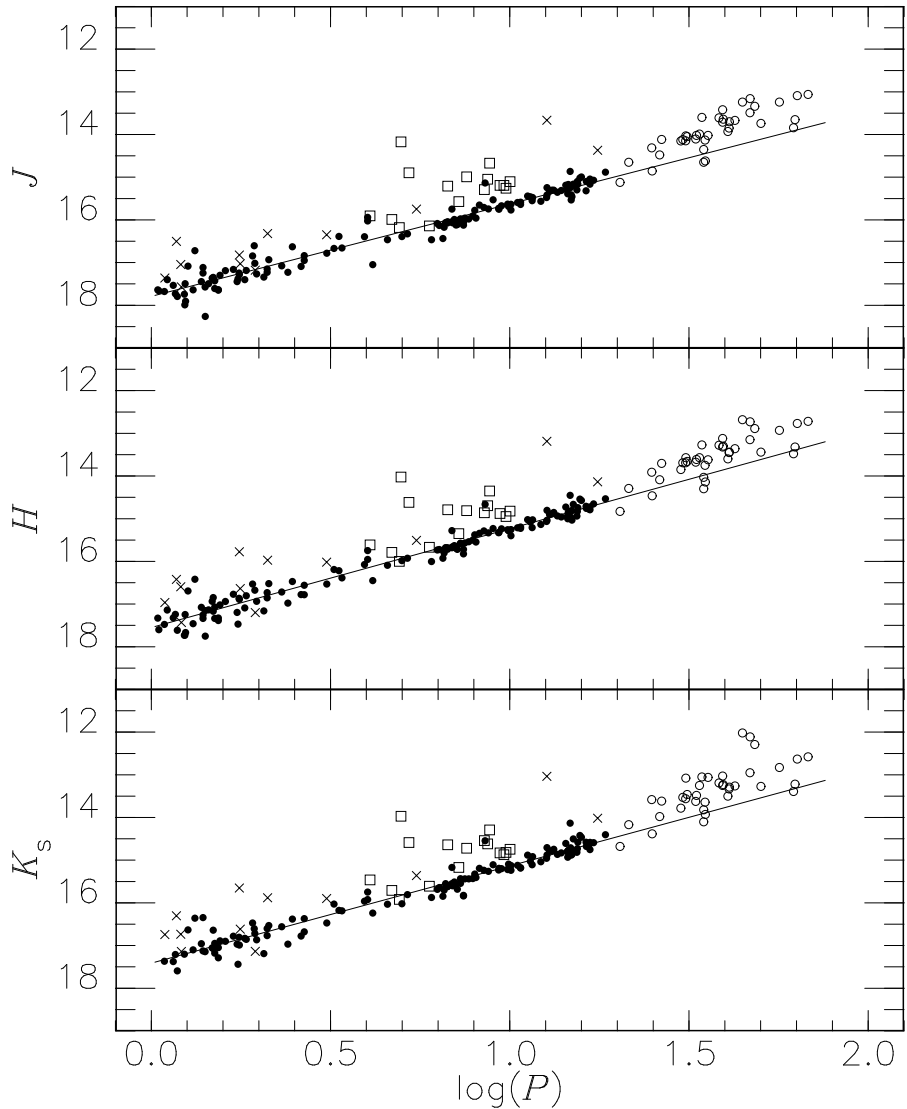}
\end{center}
\caption{Period-magnitude relation of CephIIs in the LMC.
The phase-corrected $\JHK$ magnitudes are plotted against periods.
Filled circles indicate BL+WV used to solve the PL relations
while crosses indicate those excluded,
open squares pW,
and open circles RV.
\label{fig:PLR}}
\end{minipage}
\end{figure*}

Fig.~\ref{fig:PLR} plots the phase-corrected $\JHK$ magnitudes
against $\log P$ for the LMC CephIIs.  Note that only a few of the
RVs have phase corrections applied. 
In the present section we give infrared PL relations
for the BL and WV stars using the phase-corrected data.
It is evident from Fig.~\ref{fig:PLR} that the RV stars do not
continue the linear PL relations to longer periods.  
They are discussed in section \ref{sec:RV}.

Least-square solutions yield the following relations.
For BL stars:
\begin{eqnarray}
 J_{1} = -2.164(\pm 0.240)(\log P -0.3) + 17.131(\pm 0.038), \nonumber\\
(\sigma = 0.25, 55 \: \rm {stars}) \label{eq:J1}
\end{eqnarray}
\begin{eqnarray}
 H_{1} = -2.259(\pm 0.248)(\log P -0.3) + 16.857(\pm0.039), \nonumber\\
(\sigma = 0.26, 54 \: \rm {stars}) \label{eq:H1}
\end{eqnarray}
\begin{eqnarray}
 K_{{\rm s}, 1} = -1.992(\pm 0.278)(\log P -0.3) + 16.733(\pm 0.040), \nonumber\\
(\sigma = 0.26, 47 \: \rm {stars}). \label{eq:K1}
\end{eqnarray}

For WV stars we find:
\begin{eqnarray}
J_{2} = -2.337(\pm 0.114)(\log P -1.2) + 15.165(\pm 0.030), \nonumber\\
(\sigma = 0.18, 82 \: \rm {stars}) \label{eq:J2}
\end{eqnarray}
\begin{eqnarray}
H_{2} = -2.406(\pm 0.100)(\log P -1.2) + 14.756(\pm 0.027), \nonumber\\
(\sigma = 0.16, 82 \: \rm {stars}) \label{eq:H2}
\end{eqnarray}
\begin{eqnarray}
K_{{\rm s}, 2} = -2.503(\pm 0.109)(\log P -1.2) + 14.638(\pm 0.029). \nonumber\\
(\sigma = 0.17, 82 \: \rm {stars}) \label{eq:K2}
\end{eqnarray}

The greater scatter for the BL stars compared with the WVs is at least
partly due to the poorer photometry for the fainter (BL) stars, especially
at the longer wavelengths.
Given the uncertainties in the slopes, there is no
evidence for difference between the BL and WV stars and the following are
joint solutions:
\begin{eqnarray}
J_{3} = -2.163(\pm 0.044)(\log P -1.2) + 15.194(\pm 0.029), \nonumber\\
(\sigma = 0.21, 137 \: \rm {stars}) \label{eq:J3}
\end{eqnarray}
\begin{eqnarray}
H_{3} = -2.316(\pm 0.043)(\log P -1.2) + 14.772(\pm 0.028), \nonumber\\
(\sigma = 0.20, 136 \: \rm {stars}) \label{eq:H3}
\end{eqnarray}
\begin{eqnarray}
K_{{\rm s}, 3} = -2.278(\pm 0.047)(\log P -1.2) + 14.679(\pm 0.029), \nonumber\\
(\sigma = 0.21, 129 \: \rm {stars}) \label{eq:K3}
\end{eqnarray}

The slopes do not differ significantly from those found by M06
for CephIIs in globular clusters
except possibly at $K_\mathrm{s}$ where, for instance, the slope in
eq.\  (\ref{eq:K3})
is shallower by $1.9~\sigma$ compared with the cluster value. The LMC
result is sensitive to whether we reject or retain some of the fainter
BL stars.
In the shortest period range the detection limit of the IRSF survey
can introduce a bias in the PL diagram.
In fact, all the stars with the IRSF $\Ks$ missing
are BL stars with $P$ shorter than 1.5~d.
The slope gets steeper and agrees with the case of WV (eq. \ref{eq:K2})
if we use only those with $P > 1.5$~d.
Since the difference is in any case only of marginal significance,
we have chosen to ignore it. Then, 
adopting the cluster slopes we obtain
the following relations for the combined BL and WV set:
\begin{eqnarray}
J_{4} = -2.230(\pm 0.05)(\log P -1.2) + 15.160(\pm 0.018), \nonumber\\
(\sigma = 0.21) \label{eq:J4}
\end{eqnarray}
\begin{eqnarray}
H_{4} = -2.340(\pm 0.05)(\log P -1.2) + 14.760(\pm 0.017), \nonumber\\
(\sigma = 0.20) \label{eq:H4}
\end{eqnarray}
\begin{eqnarray}
K_{{\rm s}, 4} = -2.410(\pm 0.05)(\log P -1.2) + 14.617(\pm 0.015), \nonumber\\
(\sigma = 0.21) \label{eq:K4}
\end{eqnarray}
where the uncertainties in the slopes are those given by M06.

Solutions have also been made using the LMC $\JHK$ data uncorrected for phase.
In the case of the WV stars the scatters ($\sigma$) about equations 
analogous to (\ref{eq:J2}), (\ref{eq:H2}), and (\ref{eq:K2})
are 0.26, 0.23, and 0.24~mag, respectively. This may be compared with the
figures for the phase-corrected  data; 0.18, 0.16, and 0.17~mag.
Thus a significant
improvement is found using the phase-corrected data and the scatter becomes
comparable with that found by M06 for CephII in globular clusters
(0.16, 0.15, and 0.14) even though the LMC results depend on a single
observation
per star and no allowance is made for possible differential reddening.
A similar test in the case of the BL stars shows little decrease in
the scatter
when the phase-corrected 
rather than uncorrected data are used. This is probably due to the 
poorer quality of the photometry for these fainter stars
as already noted.

\section{PL Calibration and Applications}
\label{sec:Application}
\subsection{The LMC\label{sec:muLMC}} 
  The various PL relations discussed above can be calibrated using the
two Galactic BL stars with pulsation parallaxes V553~Cen 
($\log P =0.314$) and SW~Tau ($\log P = 0.200$) 
(see F08). The intrinsic colours of these stars are $(V-I)_{0}$ = 0.70 and 0.36.
This corresponds to 0.80 and 0.46 if we redden them by an amount expected
in the LMC, where we adopt 
as a typical mean reddening, $E_{B-V} = 0.074$
(see \citealt{Caldwell-1985}). Thus
these two star span almost the whole range suggested above for the
width of the BL instability strip. 

Using the data of tables 4 and 5 of F08, we can calibrate the zero-points
of the various PL relations and hence estimate the modulus of the LMC
($\muLMC$). In
doing this the $JHK$ data of F08 were 
converted from the SAAO system into that of the IRSF using the
transformations from SAAO to 2MASS (Carpenter 2001 and web site updates)
and those from 2MASS to IRSF using the relations of Kato et al (2007).
We obtain $\muLMC$ estimates between 18.41 and 18.51
based on different PL relations obtained above
as listed in Table \ref{tab:muLMC}.

\begin{table}
\begin{center}
\begin{minipage}{60mm}
\caption{
Estimates of the distance modulus of the LMC ($\muLMC$)
based on pulsation parallaxes of V553~Cen and SW~Tau
combined with the PL relations in this paper.
\label{tab:muLMC}}
\begin{center}
\begin{tabular}{ccc}
\hline
Eq. & $\muLMC$ & Note \\
\hline
(\ref{eq:W1}) & 18.45 & $W_{1} (R=1.55)$ \\
(\ref{eq:W2}) & 18.45 & $W_{2} (R=1.45)$ \\
\hline
(\ref{eq:J1}) & 18.47 & $J_{1}$ \\
(\ref{eq:H1}) & 18.42 & $H_{1}$ \\
(\ref{eq:K1}) & 18.41 & $K_{{\rm s},1}$ \\
(\ref{eq:J1})--(\ref{eq:K1})&18.44 & mean ($J_{1}$, $H_{1}$, $K_{{\rm s},1})$\\
\hline
(\ref{eq:J4}) & 18.51 & $J_{4}$ \\
(\ref{eq:H4}) & 18.44 & $H_{4}$ \\
(\ref{eq:K4}) & 18.48 & $K_{{\rm s},4}$ \\
(\ref{eq:J4})--(\ref{eq:K4})&18.48 & mean ($J_{4}$, $H_{4}$, $K_{{\rm s},4})$\\
\hline
\end{tabular}
\end{center}
\end{minipage}
\end{center}
\end{table}

The uncertainties to be associated with these values are of interest.
The uncertainties estimated for the distance moduli of V553~Cen and SW~Tau
 are each 0.08 (F08). This thus contributes 0.06 to the uncertainties
in the above mean results. In fact the $W$ zero-points derived from
the two stars differ by 0.2~mag, i.e. an uncertainty in the mean of 0.10.
On the other hand the
two stars give mean zero-points for the $\JHK$ relations with formal
uncertainties
of only $\sim 0.04$~mag. These figures of course have their own, considerable,
uncertainty. However, remembering the range in intrinsic $(V-I)$ of the two stars,
they suggest very narrow $\JHK$ PL relations. The scatter quoted above for the
infrared PL relations must be largely due to observational effects, especially
in the case of the BL star.

Taken together these results suggest $\muLMC$ of 18.46~mag. 
They also suggest an uncertainty of less than 0.10~mag but to be
reasonably conservative we adopt this as the standard error. 
No account is taken in this of any possible metallicity effects. These values 
agree
well with those derived from classical Cepheids with trigonometrical parallaxes
by 
\citet{vanLeeuwen-2007} and \citet{Benedict-2007}.
These authors found $18.52\pm 0.03$ from a PL($W$) relation
and $18.47\pm 0.03$ from a PL($\Ks$) relation again without
metallicity corrections.
In the case of the PL($W$) relation this gave a modulus of $18.39\pm 0.05$
when a recent metallicity correction \citep{Macri-2006} 
was applied, though this correction has been questioned \citep{Bono-2008}.
These results suggest that any metallicity correction to the PL($W$) result
for CephIIs will be small. M06 found no evidence for a metallicity correction
to the $\JHK$ PL relations based on globular clusters.
Current theoretical work
(Bono, Caputo, \& Santolamazza, 1997; 
\citealt{DiCriscienzo-2007}) suggests that
the evolutionary and pulsational properties of CephIIs are only
minimally affected by metallicity.

\subsection{The distance to the Galactic centre\label{sec:muGC}}

The distance to the Galactic centre ($R_{0}$) based on the CephIIs in the
Galactic bulge was discussed in F08 on the basis of the parallaxes of
V553~Cen and SW~Tau and the $K$-band observations of
Groenewegen, Udalski \& Bono (2008).
A value of $R_{0} = 7.64 \pm 0.21$~kpc was obtained. The quoted standard error
does not take into account any systematic error in the corrections for
reddening adopted by \citet{Groenewegen-2008}.
These corrections are substantial,
ranging from 0.16 to 0.69~mag at $K$, and this may be
a significant uncertainty
in the method. It is possible in principle to use reddening-free relations.
\citeauthor{Groenewegen-2008}
illustrated this using the colour, $(I-K)$. However, in view
of the high reddening of the stars used, the result is very sensitive to
the coefficient used in the reddening-free relation (equivalent to $R$ in
section \ref{sec:optPLR}).
Infrared colours (e.g. $J-K$ or $H-K$) may be more
successful when the relevant data are available.

\subsection{NGC~6441\label{sec:mu6441}}
NGC~6441 is an outstanding example of a 'second-parameter' globular cluster.
It is of relatively high metal 
([Fe/H]$=-0.53$) but has an extended horizontal branch and RR Lyrae variables.
The RR Lyraes seem to be evolved and brighter than expected for their 
metallicity
(\citealt{Pritzl-2003}, hereafter P03).
The cluster also contains CephII for which P03
give $VI$ photometry. The reddening is large, $E_{B-V} = 0.51$ according to
P03, and possibly uncertain. This is therefore a useful cluster to
study with a reddening-free parameter. A solution of the data for the seven
CephIIs, taking $W=I -1.45(V-I)$ is,
\begin{equation}
W = -2.549(\pm 0.040) \log P  + 14.439(\pm 0.044)
\end{equation}
The slope is not significantly different from that found for the LMC
CephIIs. Calibrating this relation using the data on V553~Cen and SW~Tau
leads to a cluster modulus of $15.46\pm 0.11$~mag.
This is very close to the value
(15.48) adopted by P03 on the assumptions that the RR Lyraes in this 
metal-rich cluster have absolute magnitudes similar to those with
$\rm [Fe/H] \sim -2.0$ and that $E_{B-V} = 0.51$. Note however that the
visual absolute magnitudes of the RR Lyraes depend critically
on the reddening adopted. 

M06 list mean $\JHK$ values for two
CephIIs in NGC~6441. 
However one of these (V129) may be affected by blending (see section 5.1)
and we omit this star.
The data on the other star (V6) can be used together with the PL relations
in M06 and the data on V553~Cen and SW~Tau to obtain another
estimate of the modulus. Adopting the same reddening as above we obtain
a modulus of $15.49\pm 0.05$ consistent with 
the value just given from PL($W$). This infrared modulus also
depends on the adopted reddening. However this cannot be made much smaller
if the intrinsic colours
of the NGC~6441 CephIIs, $(V-I)_{0}$, are similar to those of the
LMC CephIIs.

\subsection{$\kappa$~Pavonis\label{sec:kappaPav}}

 $\kappa$~Pav ($\log P = 0.959$) has long been thought of as probably the
nearest CephII and hence a prime candidate for fixing the distance scale
for these objects. However, the results given in F08 were puzzling.
An apparently good pulsation parallax with a small error was derived leading
to a distance modulus of $6.55 \pm 0.07$. The revised Hipparcos parallax
led to a modulus of $5.93 \pm 0.26$;
a $2\sigma$ difference from the pulsation
parallax. It was recognized in F08 that the pulsation
parallax implied that the star was brighter than expected for a CephII.
If we adopt a LMC modulus of 18.5, the star is 0.40~mag brighter than the
LMC PL($W$) relation at its period, 0.52~mag  brighter than the PL($J$)
relation
and 0.43~mag brighter than the PL($\Ks$) relation.
If the LMC is closer than 18.5~mag, as suggested above,
these differences will be increased. 

Comparison of the data for $\kappa$~Pav
(table 5 of F08) with fig.~1 of S08 shows that in $V$ and $I$ as well as $W$ the
star lies above the normal CephII stars and in the region of the pW stars
which was identified by S08.
Further evidence that $\kappa$~Pav belongs to the pW
class is given by its colour. Its intrinsic colour is $(V-I)_{0} = 0.66$,
hence $(V-I)=0.76$ if the star is reddened
by an amount comparable with the LMC stars. Comparison
with the data of S08 (e.g.\  their fig.~2) shows that these colours are bluer
than normal CephIIs of this period (WV stars) and fall in the region
occupied by the  pW stars. 

There are two further indications that
$\kappa$~Pav belongs to the pW class. The revised Hipparcos data discussed
in F08 indicate that the star probably had a close companion. This is 
consistent with the fact that S08 find a high proportion, and probably all,
pW variables are binaries. Finally, S08 pick out pW stars based on their light
curves, the rising branch being steeper than the declining one. Comparison
of the light curve of $\kappa$~Pav = HIP~093015 (\citealt{ESA-1997},
Vol 12) with the
examples in S08 fig.~5 strongly suggests that it should be classed with the
pW variables.

\section{The Infrared Colours of type II Cepheids}

\begin{figure*}
\begin{center}
\begin{tabular}{cc}
\begin{minipage}{70mm}
\begin{center}
\includegraphics[clip,width=0.95\hsize]{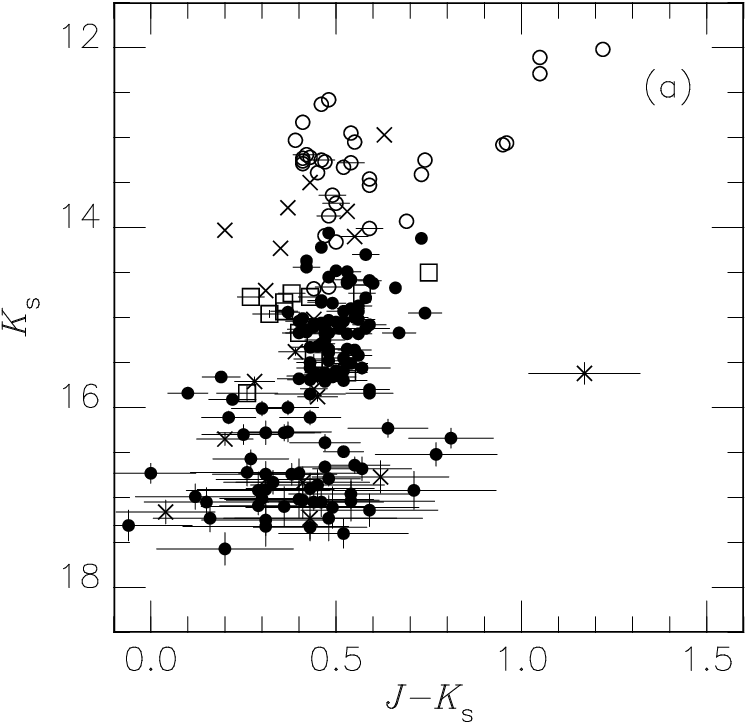}
\end{center}
\end{minipage}
\begin{minipage}{70mm}
\begin{center}
\includegraphics[clip,width=0.95\hsize]{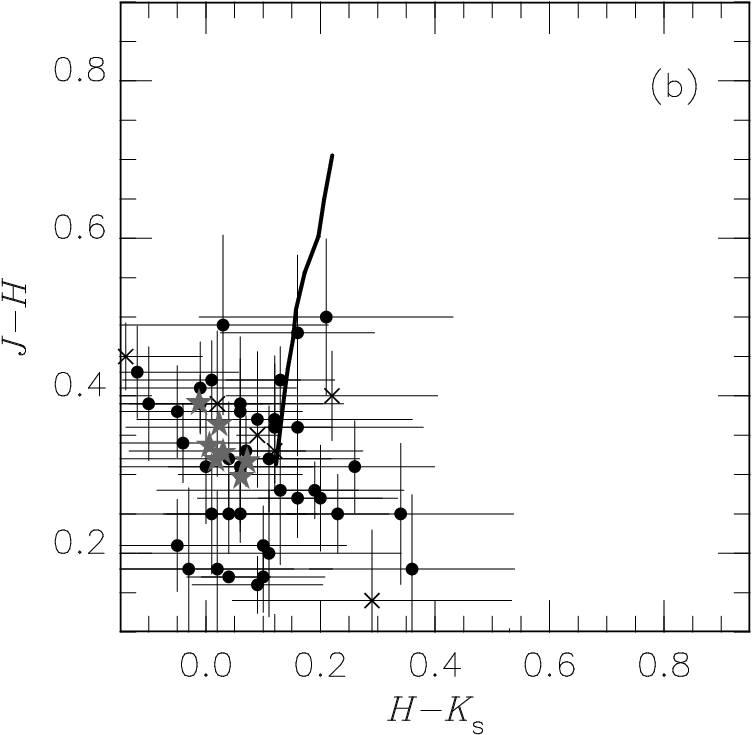}
\end{center}
\end{minipage}
\vspace{3mm}
\\
\begin{minipage}{70mm}
\begin{center}
\includegraphics[clip,width=0.95\hsize]{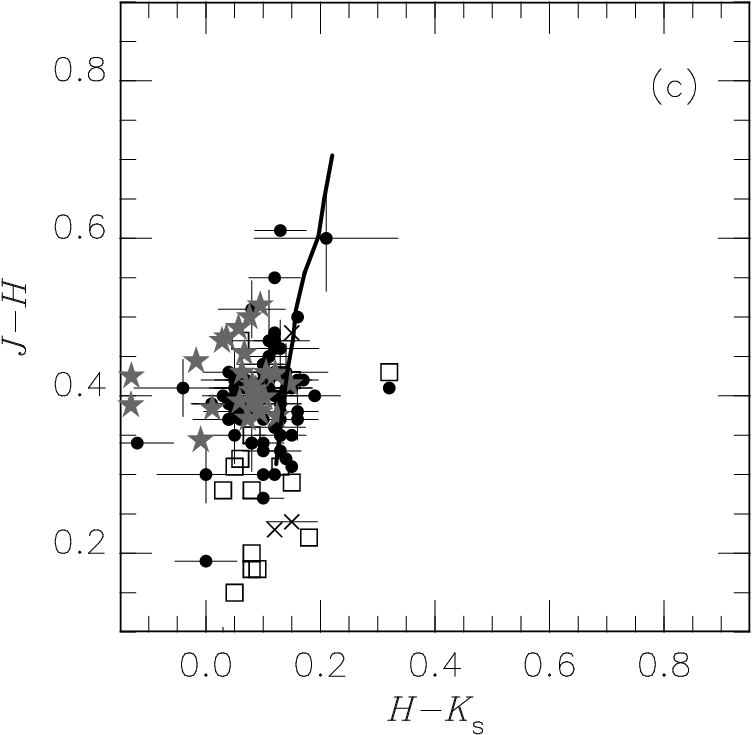}
\end{center}
\end{minipage}
\begin{minipage}{70mm}
\begin{center}
\includegraphics[clip,width=0.95\hsize]{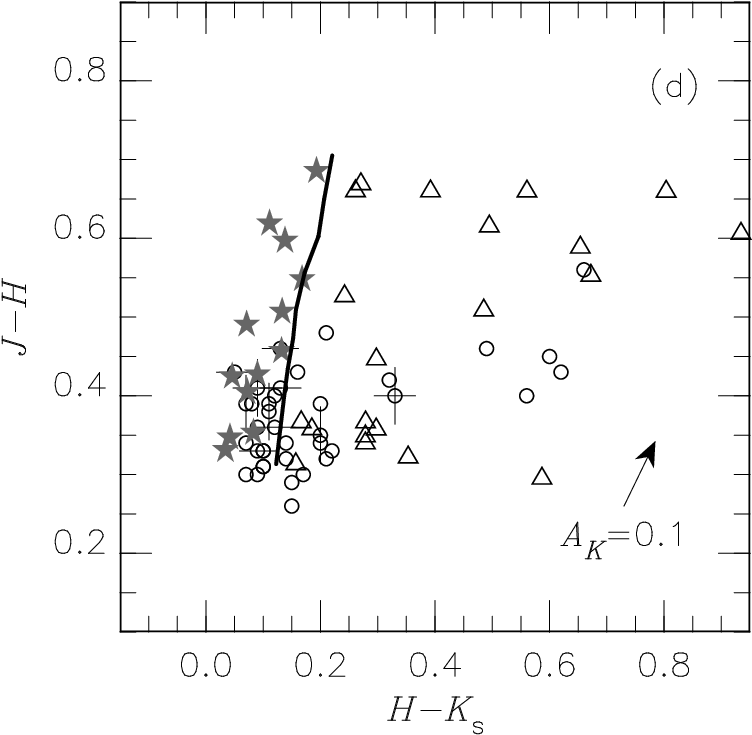}
\end{center}
\end{minipage}
\end{tabular}
\caption{
Colour-magnitude and colour-colour diagrams for CephIIs.
The panel (a) includes all the types of the LMC CephIIs,
while the panels (b), (c), and (d) are for
BL, WV+pW, and RV stars, respectively.
Symbols are the same as in Fig.~\ref{fig:PLR}
for the LMC objects, while
those in globular clusters are indicated by star symbols.
The triangles, in the panel (d), are for Galactic field RV stars
from \citet{LloydEvans-1985}.
Error bars are indicated for the LMC CephIIs
only if an uncertainty significantly exceeds the size of the symbols.
The thick curves are the loci of
local giants (see text for details).
\label{fig:CMDCCD}}
\end{center}
\end{figure*}

In this section we discuss and compare the infrared colours
of CephIIs in the LMC and those in globular
clusters (section \ref{sec:BLWV} for BL and WV, and \ref{sec:RV} for RV).
This section is essentially descriptive and further data are required
(for instance on the metallicities of the LMC stars) before detailed
interpretation of the phenomena noted can be made. 

Fig.~\ref{fig:CMDCCD} presents the colour-magnitude in the panel (a)
and colour-colour diagrams in the panels (b), (c), and (d)
for BL, WV+pW, and RV variables, respectively.
Symbols are as in Fig.~\ref{fig:PLR},
and 
the uncertainties in colour are indicated by error bars
if they exceed the size of the symbols.
We also plot CephIIs in globular clusters (M06)
in the colour-colour diagram using grey star symbols,
and 
Galactic RV stars from \citet{LloydEvans-1985} in the panel (d) with triangles.
The magnitudes of the latter sample
were converted from the SAAO system to that of the IRSF as described
in section 4.1.
No reddening corrections were applied to the LMC objects and the Galactic ones,
while, for the cluster variables,
the dereddened colours have been adjusted for the mean LMC reddening
($E_{B-V}=0.074$~mag assumed as in section \ref{sec:muLMC}),
The curve in the two-colour plots is the location of normal
giants (G0III--K5III) from \citet{Bessell-1988} whose colours are also
transformed into the IRSF/SIRIUS system after adding LMC reddening.

\begin{figure*}
\begin{minipage}{150mm}
\begin{center}
\includegraphics[clip,width=0.85\hsize]{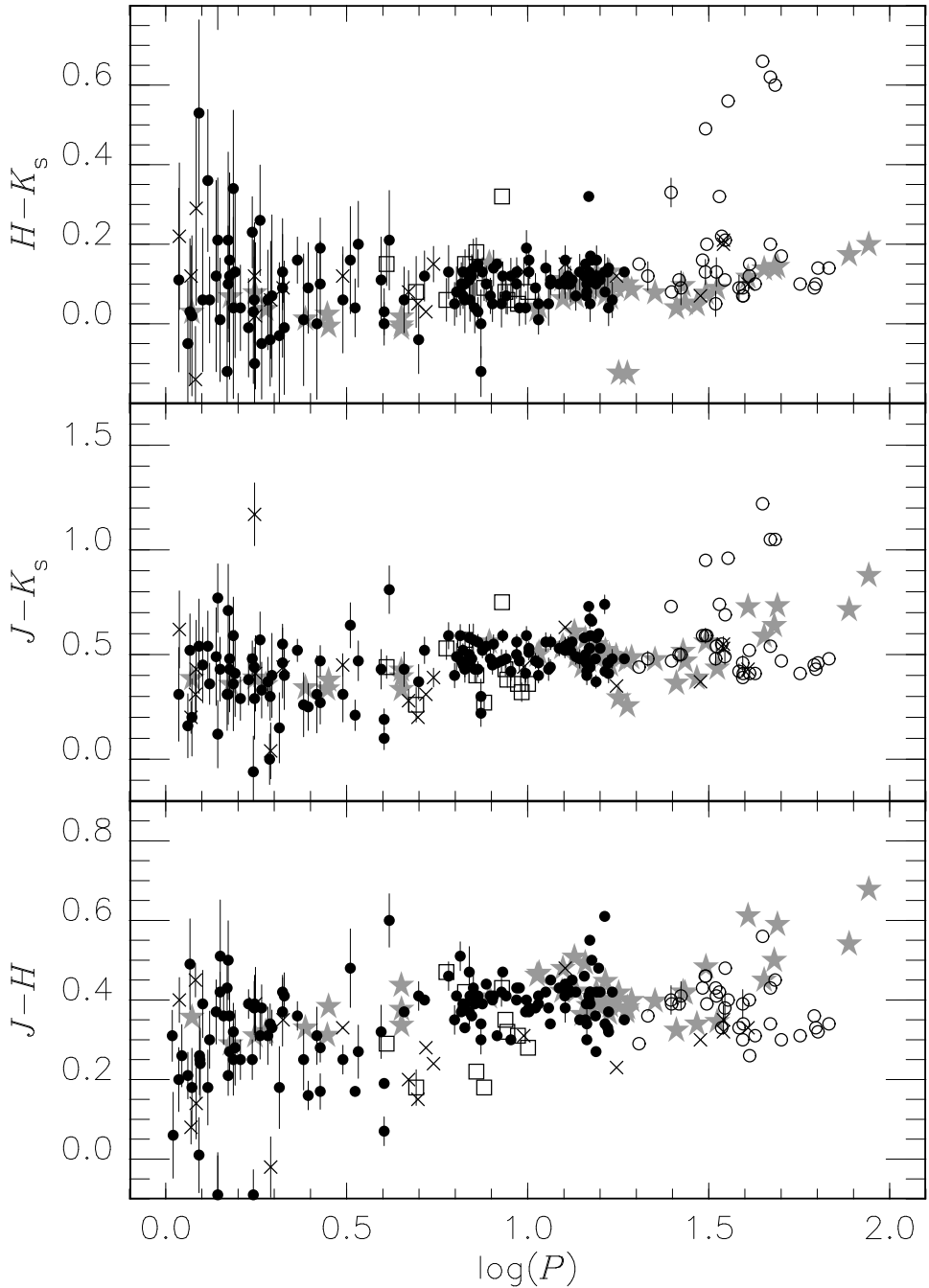}
\end{center}
\caption{
Period-colour relation.
Symbols are the same as in Fig.~\ref{fig:CMDCCD}.
Error bars are indicated for the LMC CephIIs
only if an uncertainty significantly exceeds the size of the symbols.
\label{fig:PCR}}
\end{minipage}
\end{figure*}

The infrared colour/$\log P$ relations
are plotted in Fig \ref{fig:PCR}.
Symbols are the same as in Fig.~\ref{fig:CMDCCD}
but the Galactic RVs are not included.
As in the colour-colour plots, the dereddened
magnitudes of the globular-cluster objects have been reddened by
an amount corresponding to the adopted LMC reddening,
so that the two samples are directly comparable.
Discussions on these diagrams are given below.

\subsection{BL Her and W Vir variables\label{sec:BLWV}}

Fig 3 shows that in the colour-colour plots
the cluster stars are systematically bluer in $H-\Ks$
than the local field stars (the thick line).
This can be explained, at least in a qualitative sense,
by metallicity difference;
the locus by \citet{Bessell-1988} is based on local giants
probably of the solar abundance,
while the cluster stars are metal-poor.
If we consider giants with different metallicities
but with the same surface temperature (in a range of 4000--5500~K),
a metal-poor giant is expected to be bluer than a giant with the solar abundance
in $H-\Ks$
but has almost the same colour in $J-H$ \citep{Westera-2002}.

In the case of the BL variables
the photometric uncertainties
are rather large for the LMC objects,
especially in $\Ks$, and it is difficult to discuss the distribution
of colours.
On the other hand, most of the WV stars in the LMC
form a relatively dense grouping in the panel (c) of Fig.~\ref{fig:CMDCCD},
and many of the globular-cluster stars
lie in the same region. There is a conspicuous 'line' of cluster variables
to the upper left of the main grouping. We have not found any strong 
differences between these stars and other cluster WV stars.
Two cluster variables are on the
extreme left of the Fig.~\ref{fig:CMDCCD}~(c).
One of these stars is the variable in the cluster
Terzan~1. The reddening of this cluster is high (M06 used $E_{B-V}=2.28$)
and probably rather uncertain and this makes its true position
in Fig.~\ref{fig:CMDCCD}~(c) uncertain.
The other star is V129 in NGC~6441. This star lies in a rather crowded region
of the cluster and the colours may be affected by blending.

\begin{figure}
\begin{center}
\begin{minipage}{75mm}
\begin{center}
\includegraphics[clip,width=0.95\hsize]{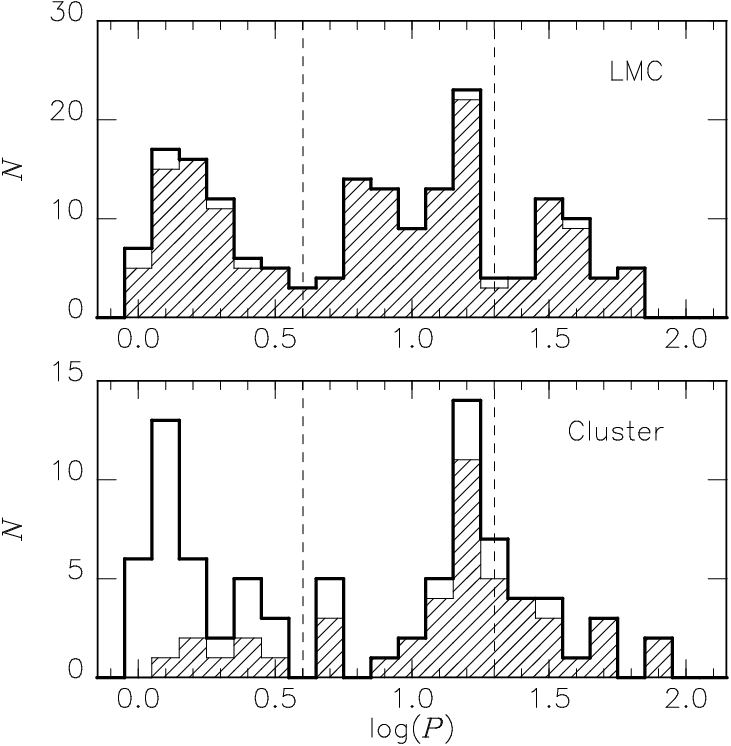}
\end{center}
\caption{
Histograms of periods for the CephIIs: the top panel
is for the LMC objects from the S08 catalogue
and the lower panel is for those in globular clusters
from a combined catalogue of M06 and P03 (see text for details).
The hatched areas indicate the period distribution of the objects
whose near-infrared magnitudes are discussed in this study.
Vertical lines indicate period
the divisions adopted by S08.
\label{fig:Phist}}
\end{minipage}
\end{center}
\end{figure}

In Fig.~\ref{fig:Phist},
we plot period histograms for CephIIs  
in the LMC (top) and
globular clusters (bottom). 
In the LMC histogram the pW stars are omitted. 
The hatched areas indicate the period distribution of the objects
whose near-infrared magnitudes are discussed in this study.
Several known CephIIs in globular clusters,
which were not observed by M06
are collected from the catalogue in \citet{Pritzl-2003}
and added as the white area in the lower panel.
A significant number of short-period BLs were not included
in M06 because of their faintness.
The previous surveys of variables with period longer than 1~d
are not complete for globular clusters.

Fig.~\ref{fig:Phist} shows a minimum in both the LMC and cluster distributions
near 4~d. This is qualitatively consistent with the division
between the BL and WV stars as summarized in
section \ref{sec:Intro}. However there are clear differences between
the two distributions at longer periods. In the range
$0.6<\log P<1.0$, there is an excess of variables in the LMC compared
with the clusters. The period distribution of the Galactic field WVs
discussed by \citet{Wallerstein-1984} seems similar to that of the
LMC field. However, a caution is required in this comparison since
various selection biases will affect the Galactic field sample including
the fact that it presumably includes Galactic field pW stars such as
$\kappa$~Pav. In the LMC the pW stars are most frequent in the range
$0.6<\log P<1.0$. It is notable that a relatively sharp peak at
$\log P\sim 1.2$ exists in both panels of Fig.~\ref{fig:Phist}.
Whilst the deficiency
of globular variables in the range $0.6<\log P< 1.0$ is in broad
qualitative agreement with Gingold's models, the excess of
variables in this period range
in the LMC field (and probably also the Galactic field) is not so easily
understood (see \citealt{Gingold-1985}, especially section 9).
However the referee has pointed out to us that the position of
the loops found by Gingold are not supported by modern models
(e.g.\  \citealt{Pietrinfenri-2006}). A closer comparison with theory
is clearly desirable but outside the scope of this paper.
The gap seen in the top panel of Fig.~\ref{fig:Phist} at
20~d, separating the LMC WV and RV stars is not present in the
cluster sample. This will be discussed in section \ref{sec:RV}.

In Fig.~\ref{fig:PCR}, WVs in the LMC and
globular-cluster groups have about the same colours at a given period.
In the case of the clusters, 
however, there is a curious tendency 
in the range $1.0 < \log P < 1.3$ for $(J-H)$  and ($J-\Ks$) to become
bluer with increasing period.
This amounts to a change in $J-H$ from $\sim 0.45$ at $\log P=1.10$
to $\sim 0.35$ at $\log P=1.25$.
Those relatively red objects at around $\log P=1.1$ belong
to the outstanding 'line' mentioned above.
Whether this feature, which is not shared by the LMC stars,
is significant, and if so what it implies, remains to be investigated.

\subsection{RV Tauri variables\label{sec:RV}}

The classical definition of an RV star is the alternating depth of
their minima. However the distinction between them and the longer-period
WV stars is not very clear. S08 adopt a division at a period (single
cycle) of 20~d. The RVs are a heterogeneous group as has long
been realized \citep{Preston-1963}
and at least some are binaries. It is clear from 
the LMC period-magnitude diagrams (Fig.~\ref{fig:PLR}, and fig.~1 of S08)
that any PL relation has a considerable scatter
though they continue the general increase of brightness with increasing
period shown by the shorter-period CephIIs.
In the case of the $\JHK$ data the plots use instantaneous
magnitudes for most of the stars, 
since phase correction is difficult for these stars; their periods
and light curves tending in many cases to be rather unstable. 

It is particularly notable that the RVs as defined by S08 do not
lie on a linear extension of the PL relations defined by shorter-period 
CephIIs ($\JHK$ and $W$ PL relations). These LMC RVs all have $\log P$ in the
range 1.3 to 1.8. As can be seen from fig.~3 of M06
CephIIs in globular clusters in this period range are co-linear with the 
shorter-period stars in $\JHK$ PL diagrams. 

Further evidence for differences between the RVs in clusters and
those in the LMC is found in the period-colour relation
(Fig.~\ref{fig:PCR}).
At the shorter periods among the RVs, $\log P \leq 1.5$,
the cluster and LMC stars tend to lie together.
They remain together
for $\log P > 1.5$ in the case of the $H-\Ks$ colours
except for the LMC objects with $\Ks$-band excess, see below,
but the situation is different in $J-\Ks$ and $J-H$.
The cluster RVs are redder in $J-H$,
while they tend to occupy a region between the bluer
LMC stars and those with an infrared excess in the case of $J-\Ks$.

The LMC RVs with the excesses include the three brightest stars
at $\Ks$ (32, 67, 174).
They are also the three reddest stars in Fig.~\ref{fig:CMDCCD},
and stand out clearly in the $\log P$-$\Ks$ diagram (Fig.~\ref{fig:PLR}).
The next two stars which are reddest in $J-\Ks$ (both with mild infrared
excesses) are numbers 91 and 180. \citet{Welch-1987} already noted numbers 67 
(HV~915) and 91 (HV~2444) 
as having infrared excesses. 
No.\  119 (HV~5829) which has a comparable excess to these two stars in 
Welch's work has the next reddest $H-\Ks$ in the IRSF data to the five stars 
just mentioned.

In the colour-colour diagram (Fig. \ref{fig:CMDCCD}d),
the LMC stars with significant $\Ks$ excesses lie
at $(H-\Ks) \sim 0.6$. Most of the others lie relatively close
to the intrinsic line, though some may have a very mild excess at $\Ks$.
Galactic RV stars from \citet{LloydEvans-1985}, triangles
in Fig.~\ref{fig:CMDCCD},
show a rather dispersed distribution; their $J-H$ colours
range from 0.1 to 0.7~mag.
No reddening correction was applied to their colours,
and some of the stars may be affected significantly.
However, this is not likely to change the overall impression
left by the plot.
Whilst there may be some
overlap between the three populations (clusters, the LMC, and the Galaxy), in
general they form distinctly different groupings. 
Their differences
have been also noted by \citet{Russell-1998}, \citet{Zsoldos-1998}, and M06.

It should be realized
that in contrast to the LMC and globular-cluster samples we do not know
the absolute magnitudes of the Galactic sample; or indeed, whether they
all are in the same evolutionary phase.
So far as the cluster RVs are concerned,
we examined the locations of those with 
periods greater than 40~d, that is the most luminous stars of the samples,
in $\Ks$-($J-\Ks$) diagrams \citep{Matsunaga-2007}.
They all lie well down the giant branch
except possibly for NGC~1904 V8 which is near the top.
Thus these longer-period stars in clusters might well be on loops
from the AGB like the, shorter-period, WV stars.
The histograms of Fig.~\ref{fig:Phist} would be consistent with the view that
for $P > 20$~d the cluster variables are simply the long-period tail of
the distribution of WVs whereas in the LMC there is a distinct 
population at these periods.

Low-resolution spectra
covering the range $\sim3800$ to $7800$~{\AA }
of all 36 of the OGLE-III RV stars were obtained in the 
period 2009 January 16--20, except for numbers 32, 58, and 162. 
The observations were made with the Cassegrain spectrograph on the
1.9-m reflector at SAAO, Sutherland.
No.\  15
had strong $\rm C_{2}$ bands and is the only one of the sample to show 
$\rm C_{2}$.
\citet{LloydEvans-2004}
examined a sample of LMC RV stars, including those
discovered in the MACHO survey and found this star 
(Macho~47.2496.8) to be the only one
in their sample with $\rm C_{2}$ bands
(see also \citealt{Pollard-2000}).
However, this star is not particularly
outstanding in any of Fig.~\ref{fig:CMDCCD} and \ref{fig:PCR}.
Note that there must be some overlap between
the two samples although \citet{LloydEvans-2004} do not list the stars they 
studied. Our
spectra show the presence of TiO in numbers 45, 75, 104, 125, 135, 149, 169
and also probably in 25, 51, 112. Since the TiO bands are known to vary
in strength with phase, it may well be present in other stars of the sample
at other times. \citet{LloydEvans-2004}
previously noted weak TiO in 169 = HV~12631.
The most sensitive indicator of the presence of TiO in our spectra 
is the $\alpha$ system sequence beginning near 6159~{\AA}.

\section{Conclusions}
\label{sec:Conclusion}
We obtained PL relations in 
phase-corrected $\JHK$ magnitudes for the combined set of BL and WV stars
in the LMC.
They have slopes consistent with those found previously (M06) in globular clusters.
For the WV stars, the scatter about the PL relations is significantly
reduced using the phase-corrected data. The longer-period variables (RVs)
show a significant scatter in the infrared and reddening-free ($VI$)
PL relations and as a group they are not co-linear with the shorter-period
stars. This contrasts with the situation for CephIIs in globular
clusters. It remains to be determined
whether those RV stars which do lie near an extrapolation
of the PLs from the shorter-period can be distinguished
from the others in some way, e.g.\  from radial velocity or spectral features.
Differences in infrared period-colour, colour-colour 
and period-frequency diagrams are found
for the WV stars in the LMC and in Galactic globular clusters.
In the case of CephII stars with $P > 20$~d (RV stars)
there are marked
differences between the available samples of the LMC, Galactic globular cluster
and the Galactic field and little evidence that the cluster stars
are post-AGB objects.
 A new class of CephIIs (pW) was identified by S08,
and we suggest that the bright Galactic star $\kappa$~Pav belongs in this 
class.
Such stars have not, so far, been identified in globular clusters and may 
thus be younger objects.

When the PL relations are calibrated using the pulsation
parallaxes of V553~Cen and SW~Tau we find a distance modulus for
the LMC of $18.46\pm 0.10$~mag without any metallicity correction, consistent
with other recent determinations. Applying our calibration to the CephII
stars in the second-parameter globular cluster NGC~6441 leads to a
modulus of $15.46\pm 0.11$~mag which confirms the view that the RR Lyraes in
this cluster are overluminous for their metallicity. 

\section*{Acknowledgments}

We thank the referee G.\  Bono for his comments and especially for
drawing our attention to some theoretical papers.

\label{lastpage}

\clearpage

\section*{Online material}

What follows is the full version of Table \ref{tab:catalogue}.
The presented data will be available in the online version of MNRAS.

\begin{table*}
\begin{minipage}{170mm}
\caption{
A catalogue of the S08 sources with IRSF counterparts
(the full version of Table \ref{tab:catalogue}).
Modified Julian Dates (MJD),
pulsation phase of the observations, $\JHK$ magnitudes, and their errors
are listed for each IRSF measurement as well as the OGLE-IDs,
types, and periods.
Shifts for the phase corrections obtained from
the $I$-band light curves are also listed if available.
Nine S08 sources are absent because their IRSF
counterparts were not found, and 15 S08 sources are listed twice
because they are identified with two counterparts from
neighbouring fields of the IRSF survey.
}
\begin{center}
\begin{tabular}{ccccccccccccc}
\hline
 OGLE- & Type & $\log P$ & \multicolumn{9}{c}{IRSF counterpart} & $\delta _{\phi}$ \\
 ID & & & IRSF-Field & MJD(obs) & Phase & $J$ & $E_J$ & $H$ & $E_H$ & $\Ks$ & $E_K$ & \\
\hline
4   & BLHer & 0.28240 & LMC0446-6800C & 52645.086 & 0.121 & 16.64 & 0.04 & 16.33 & 0.06 & 16.27 & 0.11 & $ 0.203$ \\
5   & RVTau & 1.52095 & LMC0447-7000G & 53037.995 & 0.161 & 13.82 & 0.02 & 13.41 & 0.02 & 13.28 & 0.03 & $ 0.201$ \\
6   & BLHer & 0.03660 & LMC0450-6720B & 53031.900 & 0.273 & 17.63 & 0.04 & 17.43 & 0.07 & 17.32 & 0.22 & $ 0.047$ \\
7   & BLHer & 0.09435 & LMC0452-6920F & 53044.963 & 0.224 & 17.50 & 0.04 & 17.25 & 0.06 &    -- &   -- & $-0.005$ \\
8   & BLHer & 0.24207 & LMC0451-7000E & 53308.910 & 0.293 & 17.25 & 0.02 & 17.34 & 0.06 & 17.31 & 0.17 & $ 0.130$ \\
9   & BLHer & 0.24584 & LMC0454-6700C & 53365.791 & 0.222 & 17.21 & 0.04 & 16.82 & 0.06 & 16.92 & 0.19 & $ 0.069$ \\
10  & BLHer & 0.17695 & LMC0456-6840I & 53062.869 & 0.681 & 17.71 & 0.04 & 17.35 & 0.07 & 17.23 & 0.25 & $-0.281$ \\
10  & BLHer & 0.17695 & LMC0452-6840G & 53049.895 & 0.049 & 17.33 & 0.03 & 17.06 & 0.04 & 16.90 & 0.17 & $ 0.278$ \\
11  & RVTau & 1.59391 & LMC0453-6740G & 52683.765 & 0.578 & 13.71 & 0.03 & 13.32 & 0.02 & 13.25 & 0.02 & -- \\
11  & RVTau & 1.59391 & LMC0454-6720A & 53017.960 & 0.091 & 13.42 & 0.01 & 13.12 & 0.01 & 13.03 & 0.02 & -- \\
12  & WVir  & 1.06374 & LMC0456-6820C & 53400.962 & 0.136 & 15.43 & 0.02 & 14.98 & 0.02 & 14.87 & 0.03 & $ 0.048$ \\
13  & WVir  & 1.06238 & LMC0455-7000G & 53309.080 & 0.798 & 15.51 & 0.02 & 15.17 & 0.02 & 15.07 & 0.03 & $ 0.039$ \\
14  & RVTau & 1.79152 & LMC0455-7000G & 53309.080 & 0.552 & 13.84 & 0.01 & 13.48 & 0.01 & 13.39 & 0.02 & -- \\
15  & RVTau & 1.75221 & LMC0457-6800I & 52675.879 & 0.376 & 13.24 & 0.01 & 12.93 & 0.02 & 12.83 & 0.02 & -- \\
16  & RVTau & 1.30740 & LMC0456-6900B & 53019.827 & 0.620 & 15.12 & 0.01 & 14.83 & 0.01 & 14.68 & 0.02 & -- \\
17  & WVir  & 1.16001 & LMC0456-6820H & 53053.855 & 0.031 & 14.98 & 0.01 & 14.63 & 0.01 & 14.48 & 0.02 & $ 0.323$ \\
18  & BLHer & 0.13975 & LMC0459-6920B & 53060.840 & 0.483 & 17.60 & 0.04 & 17.23 & 0.07 & 17.11 & 0.23 & $-0.153$ \\
19  & pWVir & 0.93826 & LMC0500-6800C & 53039.914 & 0.173 & 15.20 & 0.01 & 14.85 & 0.02 & 14.77 & 0.03 & -- \\
20  & BLHer & 0.04459 & LMC0500-6740C & 52604.984 & 0.503 & 17.60 & 0.05 & 17.34 & 0.06 &    -- &   -- & $-0.201$ \\
21  & pWVir & 0.98943 & LMC0501-7120I & 52935.117 & 0.360 & 15.46 & 0.02 & 15.15 & 0.02 & 15.02 & 0.03 & -- \\
22  & WVir  & 1.03006 & LMC0458-7040G & 53497.695 & 0.076 & 15.44 & 0.02 & 15.05 & 0.02 & 15.04 & 0.03 & $ 0.144$ \\
22  & WVir  & 1.03006 & LMC0502-7040I & 53117.759 & 0.623 & 15.75 & 0.02 & 15.34 & 0.02 & 15.29 & 0.03 & $-0.112$ \\
23  & pWVir & 0.71890 & LMC0500-6740E & 52605.016 & 0.366 & 15.01 & 0.02 & 14.73 & 0.02 & 14.70 & 0.02 & -- \\
24  & BLHer & 0.09575 & LMC0503-6940C & 53414.925 & 0.874 & 17.78 & 0.04 & 17.54 & 0.08 &    -- &   -- & $ 0.127$ \\
25  & RVTau & 1.83229 & LMC0503-6840I & 53301.999 & 0.989 & 13.06 & 0.02 & 12.72 & 0.01 & 12.58 & 0.01 & -- \\
26  & WVir  & 1.13283 & LMC0504-6820F & 53123.705 & 0.243 & 15.33 & 0.02 & 14.91 & 0.02 & 14.84 & 0.03 & $ 0.025$ \\
27  & WVir  & 1.23385 & LMC0503-6940E & 53053.791 & 0.478 & 15.51 & 0.02 & 15.09 & 0.02 & 15.03 & 0.04 & $-0.438$ \\
28  & pWVir & 0.94373 & LMC0502-7000B & 53083.751 & 0.192 & 15.11 & 0.02 & 14.79 & 0.02 & 14.73 & 0.03 & -- \\
29  & RVTau & 1.49479 & LMC0503-6840E & 53301.952 & 0.393 & 14.05 & 0.02 & 13.66 & 0.01 & 13.46 & 0.02 & -- \\
30  & BLHer & 0.59499 & LMC0504-6800B & 53118.735 & 0.170 & 16.28 & 0.03 & 15.96 & 0.06 & 15.85 & 0.09 & $ 0.113$ \\
31  & WVir  & 0.82646 & LMC0506-7120I & 52932.098 & 0.905 & 15.99 & 0.02 & 15.66 & 0.02 & 15.56 & 0.03 & $ 0.054$ \\
32  & RVTau & 1.64896 & LMC0504-6720B & 53344.945 & 0.762 & 13.24 & 0.01 & 12.68 & 0.01 & 12.02 & 0.01 & -- \\
33  & pWVir & 0.97288 & LMC0503-6900G & 53305.936 & 0.099 & 15.19 & 0.02 & 14.88 & 0.02 & 14.83 & 0.02 & -- \\
34  & WVir  & 1.17351 & LMC0503-6900A & 53302.014 & 0.084 & 15.18 & 0.01 & 14.76 & 0.01 & 14.59 & 0.02 & $ 0.274$ \\
35  & WVir  & 0.99414 & LMC0503-6920D & 53375.777 & 0.942 & 15.54 & 0.01 & 15.17 & 0.01 & 15.13 & 0.04 & $ 0.090$ \\
36  & WVir  & 1.17262 & LMC0507-6920C & 52612.966 & 0.513 & 15.54 & 0.02 & 15.15 & 0.02 & 15.10 & 0.04 & $-0.343$ \\
37  & WVir  & 0.83869 & LMC0506-7120E & 52921.052 & 0.510 & 16.09 & 0.02 & 15.72 & 0.03 & 15.59 & 0.05 & $-0.039$ \\
38  & WVir  & 0.60354 & LMC0503-6840A & 53123.751 & 0.964 & 15.94 & 0.02 & 15.87 & 0.03 & 15.84 & 0.05 & $ 0.084$ \\
38  & WVir  & 0.60354 & LMC0507-6840C & 53341.916 & 0.320 & 15.85 & 0.01 & 15.66 & 0.02 & 15.66 & 0.05 & $ 0.089$ \\
39  & WVir  & 0.94031 & LMC0504-6720G & 53354.888 & 0.426 & 15.83 & 0.01 & 15.42 & 0.02 & 15.35 & 0.04 & $-0.087$ \\
40  & pWVir & 0.98345 & LMC0506-6940C & 53308.992 & 0.829 & 15.28 & 0.02 &    -- &   -- & 14.96 & 0.04 & -- \\
41  & BLHer & 0.39377 & LMC0507-6920F & 52613.016 & 0.105 & 16.55 & 0.02 & 16.39 & 0.03 & 16.30 & 0.11 & $ 0.080$ \\
42  & pWVir & 0.69220 & LMC0507-6920I & 52613.884 & 0.120 & 16.10 & 0.02 & 15.92 & 0.04 & 15.84 & 0.08 & -- \\
43  & WVir  & 0.81687 & LMC0506-7000H & 53080.865 & 0.265 & 16.18 & 0.02 & 15.81 & 0.02 & 15.71 & 0.05 & $-0.007$ \\
44  & WVir  & 1.12287 & LMC0506-6940B & 53308.980 & 0.384 & 15.49 & 0.02 & 15.04 & 0.01 & 14.93 & 0.02 & $-0.186$ \\
45  & RVTau & 1.80200 & LMC0506-6940H & 53309.089 & 0.315 & 13.09 & 0.02 & 12.77 & 0.01 & 12.63 & 0.01 & -- \\
46  & WVir  & 1.16861 & LMC0507-6820E & 53081.785 & 0.371 & 14.85 & 0.01 & 14.44 & 0.01 & 14.12 & 0.02 & $ 0.013$ \\
47  & WVir  & 0.86250 & LMC0506-6940D & 53309.004 & 0.917 & 15.93 & 0.02 & 15.53 & 0.02 & 15.50 & 0.03 & $ 0.045$ \\
48  & BLHer & 0.16000 & LMC0507-6820A & 53078.859 & 0.689 & 17.77 & 0.04 & 17.41 & 0.06 &    -- &   -- & $-0.274$ \\
49  & BLHer & 0.50991 & LMC0510-6940I & 53333.028 & 0.613 & 16.87 & 0.04 & 16.39 & 0.09 & 16.23 & 0.10 & $-0.201$ \\
\hline
\end{tabular}
\end{center}
\end{minipage}
\end{table*}

\addtocounter{table}{-1}
\begin{table*}
\begin{minipage}{170mm}
\caption{--continued.}
\begin{center}
\begin{tabular}{ccccccccccccc}
\hline
 OGLE- & Type & $\log P$ & \multicolumn{9}{c}{IRSF counterpart} & $\delta _{\phi}$ \\
 ID & & & IRSF-Field & MJD(obs) & Phase & $J$ & $E_J$ & $H$ & $E_H$ & $\Ks$ & $E_K$ & \\
\hline
50  & RVTau & 1.54093 & LMC0511-6840C & 53051.855 & 0.304 & 14.35 & 0.02 & 14.03 & 0.01 & 13.82 & 0.02 & -- \\
50  & RVTau & 1.54093 & LMC0511-6900I & 52612.916 & 0.672 & 14.65 & 0.03 & 14.30 & 0.02 & 14.10 & 0.02 & -- \\
51  & RVTau & 1.60859 & LMC0511-6900I & 52612.916 & 0.627 & 13.93 & 0.02 & 13.60 & 0.02 & 13.50 & 0.02 & -- \\
52  & pWVir & 0.67098 & LMC0510-7000E & 53322.849 & 0.132 & 15.99 & 0.02 & 15.79 & 0.02 & 15.71 & 0.05 & -- \\
53  & BLHer & 0.01828 & LMC0511-6800E & 53051.893 & 0.747 & 17.98 & 0.04 & 17.67 & 0.05 &    -- &   -- & $-0.342$ \\
54  & WVir  & 0.99674 & LMC0510-7020D & 53136.698 & 0.196 & 15.67 & 0.02 & 15.27 & 0.02 & 15.08 & 0.04 & $ 0.013$ \\
55  & RVTau & 1.61284 & LMC0511-6800H & 53060.790 & 0.730 & 13.70 & 0.02 & 13.44 & 0.01 & 13.29 & 0.02 & -- \\
56  & WVir  & 0.86271 & LMC0510-6940G & 53333.007 & 0.167 & 15.98 & 0.02 & 15.57 & 0.02 & 15.42 & 0.03 & $ 0.013$ \\
57  & WVir  & 1.22095 & LMC0511-6840D & 53051.872 & 0.704 & 15.29 & 0.02 & 14.96 & 0.02 & 14.83 & 0.03 & $-0.187$ \\
58  & RVTau & 1.33209 & LMC0511-6840G & 53060.769 & 0.606 & 15.14 & 0.02 & 14.78 & 0.02 & 14.66 & 0.03 & $-0.490$ \\
59  & WVir  & 1.22365 & LMC0510-7040G & 53322.877 & 0.012 & 14.68 & 0.01 & 14.36 & 0.02 & 14.22 & 0.02 & $ 0.365$ \\
59  & WVir  & 1.22365 & LMC0514-7040I & 52714.762 & 0.677 & 15.42 & 0.02 & 15.05 & 0.02 & 15.01 & 0.04 & $-0.258$ \\
60  & BLHer & 0.09223 & LMC0514-7020I & 52697.744 & 0.853 & 17.58 & 0.05 & 17.57 & 0.08 & 17.04 & 0.22 & $ 0.162$ \\
61  & BLHer & 0.07244 & LMC0514-6900C & 52595.083 & 0.780 & 17.77 & 0.04 & 17.59 & 0.09 & 17.57 & 0.18 & $ 0.023$ \\
62  & WVir  & 0.78152 & LMC0514-6940F & 52593.994 & 0.109 & 16.43 & 0.02 & 15.97 & 0.03 & 15.84 & 0.06 & $ 0.033$ \\
63  & WVir  & 0.84039 & LMC0514-7000H & 52692.884 & 0.233 & 16.08 & 0.02 & 15.67 & 0.02 & 15.61 & 0.05 & $-0.011$ \\
64  & BLHer & 0.32795 & LMC0514-6840E & 52650.823 & 0.603 & 17.13 & 0.03 & 16.72 & 0.05 & 16.73 & 0.16 & $-0.198$ \\
65  & RVTau & 1.54475 & LMC0514-6900E & 52598.969 & 0.170 & 14.13 & 0.03 & 13.75 & 0.02 & 13.64 & 0.02 & -- \\
66  & WVir  & 1.11758 & LMC0515-6720H & 52305.855 & 0.288 & 15.45 & 0.02 & 15.03 & 0.02 & 14.93 & 0.04 & $-0.147$ \\
67  & RVTau & 1.68333 & LMC0514-6920H & 52594.975 & 0.629 & 13.34 & 0.01 & 12.89 & 0.01 & 12.29 & 0.02 & -- \\
68  & BLHer & 0.20664 & LMC0514-6900E & 52598.969 & 0.607 & 17.38 & 0.03 & 17.13 & 0.04 & 17.09 & 0.10 & $-0.190$ \\
70  & WVir  & 1.18859 & LMC0515-6640E & 52288.779 & 0.645 & 15.31 & 0.01 & 15.04 & 0.02 & 14.94 & 0.03 & $-0.209$ \\
71  & BLHer & 0.06151 & LMC0514-6900G & 52599.001 & 0.169 & 17.39 & 0.03 & 17.18 & 0.05 & 17.23 & 0.15 & $ 0.144$ \\
72  & WVir  & 1.16178 & LMC0515-6700D & 52338.794 & 0.897 & 15.27 & 0.02 & 14.88 & 0.01 & 14.81 & 0.03 & $ 0.120$ \\
73  & BLHer & 0.48966 & LMC0515-6720D & 53402.954 & 0.904 & 16.59 & 0.02 & 16.34 & 0.03 & 16.28 & 0.13 & $ 0.192$ \\
74  & WVir  & 0.95368 & LMC0514-6840A & 52648.819 & 0.512 & 15.58 & 0.02 & 15.28 & 0.02 & 15.16 & 0.04 & $-0.051$ \\
75  & RVTau & 1.70059 & LMC0518-6940C & 52294.792 & 0.147 & 13.74 & 0.01 & 13.44 & 0.02 & 13.27 & 0.01 & -- \\
76  & BLHer & 0.32311 & LMC0518-6820I & 52657.914 & 0.487 & 17.19 & 0.03 & 16.77 & 0.03 & 16.64 & 0.09 & $-0.039$ \\
76  & BLHer & 0.32311 & LMC0514-6820G & 52660.786 & 0.852 & 17.51 & 0.05 & 17.14 & 0.07 & 17.05 & 0.16 & $-0.284$ \\
77  & BLHer & 0.08415 & LMC0518-6940F & 52294.841 & 0.534 & 17.66 & 0.04 & 17.52 & 0.08 & 17.23 & 0.23 & $-0.090$ \\
78  & pWVir & 0.82713 & LMC0518-6920C & 52303.770 & 0.294 & 15.30 & 0.02 & 14.88 & 0.02 & 14.73 & 0.03 & -- \\
79  & WVir  & 1.17159 & LMC0518-6820F & 52657.863 & 0.117 & 15.13 & 0.02 & 14.71 & 0.02 & 14.60 & 0.02 & $ 0.283$ \\
80  & RVTau & 1.61190 & LMC0518-6940C & 52294.792 & 0.431 & 13.85 & 0.01 & 13.45 & 0.01 & 13.33 & 0.01 & -- \\
81  & WVir  & 0.97679 & LMC0518-7140C & 52727.774 & 0.084 & 15.60 & 0.02 & 15.17 & 0.02 & 15.13 & 0.04 & $ 0.069$ \\
82  & RVTau & 1.54561 & LMC0518-7140F & 53356.035 & 0.488 & 14.62 & 0.01 & 14.14 & 0.01 & 13.93 & 0.02 & -- \\
83  & pWVir & 0.77580 & LMC0518-7000I & 52288.843 & 0.434 & 16.14 & 0.02 & 15.67 & 0.02 & 15.61 & 0.05 & -- \\
84  & BLHer & 0.24818 & LMC0518-6920C & 52303.770 & 0.675 & 17.24 & 0.06 & 16.85 & 0.07 & 16.83 & 0.17 & $-0.214$ \\
85  & BLHer & 0.53213 & LMC0518-7120E & 52761.750 & 0.704 & 16.86 & 0.03 & 16.59 & 0.06 & 16.39 & 0.09 & $-0.202$ \\
86  & WVir  & 1.19991 & LMC0518-6940E & 52294.825 & 0.278 & 15.15 & 0.02 & 14.73 & 0.01 & 14.62 & 0.02 & $-0.156$ \\
87  & WVir  & 0.71475 & LMC0518-6940E & 52294.825 & 0.118 & 16.22 & 0.02 & 15.82 & 0.02 & 15.70 & 0.06 & $ 0.106$ \\
88  & BLHer & 0.29020 & LMC0518-7100H & 52754.751 & 0.503 & 17.20 & 0.03 & 17.22 & 0.07 & 17.16 & 0.13 & $-0.022$ \\
89  & BLHer & 0.06718 & LMC0518-6940B & 53332.028 & 0.561 & 17.92 & 0.07 & 17.43 & 0.09 & 17.40 & 0.16 & $-0.189$ \\
90  & BLHer & 0.16995 & LMC0518-6800E & 52661.030 & 0.633 & 17.56 & 0.03 & 17.13 & 0.05 & 17.25 & 0.17 & $-0.195$ \\
91  & RVTau & 1.55327 & LMC0518-6900D & 52590.947 & 0.590 & 14.02 & 0.02 & 13.62 & 0.01 & 13.06 & 0.02 & -- \\
92  & BLHer & 0.41777 & LMC0518-7000D & 52288.914 & 0.702 & 17.22 & 0.04 & 16.91 & 0.06 & 16.91 & 0.18 & $-0.131$ \\
93  & WVir  & 1.24534 & LMC0518-7000G & 52288.809 & 0.771 & 14.58 & 0.02 & 14.35 & 0.02 & 14.23 & 0.02 & $-0.212$ \\
94  & WVir  & 0.92781 & LMC0522-7000I & 53421.909 & 0.708 & 15.72 & 0.02 & 15.30 & 0.03 & 15.25 & 0.05 & $-0.012$ \\
95  & WVir  & 0.69898 & LMC0522-6820F & 52654.900 & 0.829 & 16.37 & 0.02 & 15.96 & 0.03 & 16.00 & 0.08 & $ 0.022$ \\
96  & WVir  & 1.14382 & LMC0522-6840C & 52637.927 & 0.566 & 15.65 & 0.02 & 15.30 & 0.02 & 15.17 & 0.04 & $-0.339$ \\
97  & WVir  & 1.02161 & LMC0522-6920I & 52301.918 & 0.620 & 15.65 & 0.02 & 15.27 & 0.02 & 15.18 & 0.03 & $-0.057$ \\
\hline
\end{tabular}
\end{center}
\end{minipage}
\end{table*}

\addtocounter{table}{-1}
\begin{table*}
\begin{minipage}{170mm}
\caption{--continued.}
\begin{center}
\begin{tabular}{ccccccccccccc}
\hline
 OGLE- & Type & $\log P$ & \multicolumn{9}{c}{IRSF counterpart} & $\delta _{\phi}$ \\
 ID & & & IRSF-Field & MJD(obs) & Phase & $J$ & $E_J$ & $H$ & $E_H$ & $\Ks$ & $E_K$ & \\
\hline
98  & pWVir & 0.69668 & LMC0522-7020I & 52702.739 & 0.734 & 14.23 & 0.02 & 14.08 & 0.01 & 14.03 & 0.02 & -- \\
99  & WVir  & 1.18996 & LMC0522-6900F & 52637.806 & 0.002 & 14.88 & 0.02 & 14.46 & 0.02 & 14.30 & 0.03 & $ 0.280$ \\
100 & WVir  & 0.87105 & LMC0522-7020F & 53355.891 & 0.554 & 16.31 & 0.02 & 16.01 & 0.03 & 16.01 & 0.08 & $-0.187$ \\
100 & WVir  & 0.87105 & LMC0522-7020E & 52700.891 & 0.411 & 16.13 & 0.02 & 15.79 & 0.02 & 15.91 & 0.06 & $-0.072$ \\
101 & WVir  & 1.05761 & LMC0522-6920H & 52301.904 & 0.676 & 15.76 & 0.02 & 15.38 & 0.02 & 15.33 & 0.05 & $-0.302$ \\
102 & BLHer & 0.10244 & LMC0522-7000E & 52289.769 & 0.582 & 17.31 & 0.06 & 16.92 & 0.06 & 16.86 & 0.17 & $-0.226$ \\
103 & WVir  & 1.11087 & LMC0522-7020H & 52701.758 & 0.478 & 15.55 & 0.02 & 15.12 & 0.02 & 15.00 & 0.03 & $-0.218$ \\
104 & RVTau & 1.39585 & LMC0522-7000B & 52287.914 & 0.982 & 14.14 & 0.02 & 13.74 & 0.03 & 13.41 & 0.02 & $ 0.172$ \\
105 & BLHer & 0.17298 & LMC0522-7020E & 52700.891 & 0.040 & 17.05 & 0.03 & 16.84 & 0.04 & 16.74 & 0.14 & $ 0.328$ \\
105 & BLHer & 0.17298 & LMC0522-7020H & 52701.758 & 0.622 & 17.63 & 0.07 & 17.13 & 0.07 & 16.92 & 0.21 & $-0.281$ \\
106 & WVir  & 0.82651 & LMC0522-6920B & 52296.810 & 0.608 & 16.10 & 0.02 & 15.71 & 0.02 & 15.65 & 0.08 & $-0.029$ \\
107 & BLHer & 0.08248 & LMC0522-6940E & 52293.857 & 0.730 & 17.11 & 0.03 & 16.66 & 0.03 & 16.80 & 0.13 & $-0.066$ \\
108 & RVTau & 1.47728 & LMC0522-6820H & 52655.794 & 0.076 & 14.15 & 0.02 & 13.85 & 0.02 & 13.78 & 0.02 & -- \\
109 & BLHer & 0.15062 & LMC0522-6940E & 52293.857 & 0.525 & 18.44 & 0.10 & 17.93 & 0.10 &    -- &   -- & $-0.179$ \\
110 & WVir  & 0.84994 & LMC0522-6900H & 52637.858 & 0.722 & 16.13 & 0.03 & 15.70 & 0.02 & 15.56 & 0.07 & $-0.010$ \\
111 & WVir  & 0.87481 & LMC0522-7100H & 52760.825 & 0.717 & 15.97 & 0.02 & 15.58 & 0.02 & 15.45 & 0.03 & $-0.011$ \\
112 & RVTau & 1.59547 & LMC0522-6920A & 52296.794 & 0.608 & 13.64 & 0.01 & 13.30 & 0.01 & 13.23 & 0.02 & -- \\
113 & BLHer & 0.48932 & LMC0522-6940G & 52293.893 & 0.145 & 16.33 & 0.02 & 16.00 & 0.02 & 15.88 & 0.05 & $ 0.018$ \\
114 & BLHer & 0.03786 & LMC0525-6820F & 52681.761 & 0.833 & 17.39 & 0.04 & 16.99 & 0.04 & 16.77 & 0.18 & $-0.027$ \\
115 & RVTau & 1.39736 & LMC0522-6940G & 52293.893 & 0.950 & 14.56 & 0.01 & 14.17 & 0.01 & 14.09 & 0.02 & $ 0.293$ \\
116 & BLHer & 0.29373 & LMC0526-6920C & 52294.942 & 0.585 & 17.42 & 0.02 & 17.09 & 0.04 & 17.02 & 0.20 & $-0.154$ \\
117 & WVir  & 0.82147 & LMC0526-7100C & 52939.119 & 0.767 & 16.08 & 0.02 & 15.70 & 0.01 & 15.63 & 0.03 & $-0.017$ \\
118 & WVir  & 1.10376 & LMC0525-6800B & 52673.837 & 0.240 & 15.44 & 0.02 & 15.00 & 0.02 & 14.90 & 0.03 & $-0.195$ \\
118 & WVir  & 1.10376 & LMC0525-6820H & 52683.781 & 0.023 & 15.12 & 0.03 & 14.74 & 0.02 & 14.58 & 0.03 & $ 0.297$ \\
119 & RVTau & 1.52924 & LMC0526-7100I & 52951.020 & 0.442 & 13.99 & 0.01 & 13.57 & 0.01 & 13.25 & 0.02 & -- \\
120 & WVir  & 0.65887 & LMC0525-6840B & 52683.843 & 0.753 & 16.54 & 0.02 & 16.17 & 0.02 & 16.11 & 0.08 & $-0.077$ \\
121 & BLHer & 0.31416 & LMC0526-7020E & 52773.712 & 0.871 & 17.20 & 0.05 & 17.02 & 0.09 & 17.05 & 0.16 & $ 0.140$ \\
122 & BLHer & 0.18715 & LMC0525-6840H & 52683.954 & 0.007 & 17.46 & 0.04 & 17.14 & 0.06 & 17.10 & 0.22 & $ 0.186$ \\
122 & BLHer & 0.18715 & LMC0525-6820B & 52675.910 & 0.779 & 17.73 & 0.04 & 17.48 & 0.08 & 17.14 & 0.18 & $-0.093$ \\
124 & BLHer & 0.23927 & LMC0525-6900G & 53333.071 & 0.059 & 17.27 & 0.03 & 17.02 & 0.04 & 16.79 & 0.08 & $ 0.176$ \\
125 & RVTau & 1.51896 & LMC0526-7140A & 52959.955 & 0.407 & 14.35 & 0.01 & 13.92 & 0.01 & 13.87 & 0.03 & $-0.245$ \\
126 & WVir  & 1.21290 & LMC0526-7100G & 53364.763 & 0.361 & 15.69 & 0.02 & 15.08 & 0.02 & 14.95 & 0.04 & $-0.370$ \\
127 & WVir  & 1.10275 & LMC0529-6920C & 52362.833 & 0.694 & 15.56 & 0.02 & 15.15 & 0.02 & 15.04 & 0.03 & $-0.084$ \\
128 & WVir  & 1.26700 & LMC0530-7020I & 52727.723 & 0.022 & 14.54 & 0.01 & 14.19 & 0.01 & 14.06 & 0.02 & $ 0.343$ \\
129 & RVTau & 1.79594 & LMC0530-7000I & 52338.892 & 0.062 & 13.65 & 0.02 & 13.32 & 0.01 & 13.22 & 0.02 & -- \\
130 & BLHer & 0.28885 & LMC0530-7040F & 52921.131 & 0.746 & 17.31 & 0.03 & 16.97 & 0.04 & 17.01 & 0.11 & $-0.293$ \\
131 & BLHer & 0.15029 & LMC0529-6840H & 52685.851 & 0.522 & 17.76 & 0.03 & 17.34 & 0.04 & 17.33 & 0.15 & $-0.186$ \\
132 & pWVir & 1.00077 & LMC0530-7000H & 52338.874 & 0.779 & 15.29 & 0.02 & 15.01 & 0.02 & 14.93 & 0.03 & -- \\
133 & WVir  & 0.79809 & LMC0530-7020B & 52725.792 & 0.718 & 16.08 & 0.03 & 15.73 & 0.02 & 15.68 & 0.05 & $ 0.004$ \\
134 & pWVir & 0.61020 & LMC0530-6940B & 52338.925 & 0.466 & 15.90 & 0.02 & 15.61 & 0.02 & 15.46 & 0.05 & -- \\
135 & RVTau & 1.42361 & LMC0529-6920H & 52363.787 & 0.294 & 14.23 & 0.02 & 13.82 & 0.03 & 13.73 & 0.03 & $-0.115$ \\
136 & BLHer & 0.12157 & LMC0530-6940H & 52362.759 & 0.805 & 16.64 & 0.04 & 16.34 & 0.06 & 16.28 & 0.09 & $ 0.080$ \\
137 & WVir  & 0.80362 & LMC0530-6940E & 52338.977 & 0.355 & 16.15 & 0.02 & 15.74 & 0.02 & 15.66 & 0.06 & $-0.019$ \\
138 & BLHer & 0.14414 & LMC0529-6840A & 52684.777 & 0.524 & 17.29 & 0.10 & 17.43 & 0.11 & 16.52 & 0.13 & $-0.175$ \\
138 & BLHer & 0.14414 & LMC0529-6900G & 52364.819 & 0.931 & 17.11 & 0.08 & 17.20 & 0.07 & 16.99 & 0.14 & $ 0.135$ \\
139 & WVir  & 1.16969 & LMC0529-6900A & 52363.822 & 0.081 & 15.02 & 0.02 & 14.62 & 0.01 & 14.49 & 0.03 & $ 0.220$ \\
140 & BLHer & 0.26509 & LMC0529-6920G & 52363.769 & 0.331 & 17.16 & 0.03 & 16.78 & 0.05 & 16.83 & 0.13 & $ 0.025$ \\
141 & BLHer & 0.26078 & LMC0531-7140E & 52960.998 & 0.069 & 17.25 & 0.03 & 16.94 & 0.05 & 16.68 & 0.13 & $ 0.149$ \\
142 & BLHer & 0.24570 & LMC0532-6800C & 52765.718 & 0.313 & 16.79 & 0.09 & 15.74 & 0.06 & 15.62 & 0.12 & $ 0.034$ \\
143 & WVir  & 1.16347 & LMC0529-6920G & 52363.769 & 0.569 & 15.57 & 0.02 & 15.27 & 0.02 & 15.17 & 0.04 & $-0.401$ \\
\hline
\end{tabular}
\end{center}
\end{minipage}
\end{table*}

\addtocounter{table}{-1}
\begin{table*}
\begin{minipage}{170mm}
\caption{--continued.}
\begin{center}
\begin{tabular}{ccccccccccccc}
\hline
 OGLE- & Type & $\log P$ & \multicolumn{9}{c}{IRSF counterpart} & $\delta _{\phi}$ \\
 ID & & & IRSF-Field & MJD(obs) & Phase & $J$ & $E_J$ & $H$ & $E_H$ & $\Ks$ & $E_K$ & \\
\hline
143 & WVir  & 1.16347 & LMC0533-6920I & 53331.899 & 0.014 & 14.86 & 0.03 & 14.52 & 0.02 & 14.44 & 0.02 & $ 0.332$ \\
144 & BLHer & 0.28723 & LMC0533-6900I & 52286.935 & 0.371 & 16.73 & 0.05 &    -- &   -- & 16.73 & 0.11 & $-0.124$ \\
145 & BLHer & 0.52340 & LMC0533-6900F & 52283.877 & 0.087 & 16.32 & 0.02 & 16.15 & 0.02 & 16.11 & 0.07 & $ 0.067$ \\
146 & WVir  & 1.00344 & LMC0533-6840C & 52687.758 & 0.228 & 15.63 & 0.02 & 15.25 & 0.01 & 15.12 & 0.03 & $ 0.006$ \\
146 & WVir  & 1.00344 & LMC0533-6900I & 52286.935 & 0.463 & 15.88 & 0.01 & 15.51 & 0.02 & 15.35 & 0.03 & $-0.111$ \\
147 & RVTau & 1.67021 & LMC0533-6920I & 53331.899 & 0.585 & 13.49 & 0.02 & 13.15 & 0.01 & 12.95 & 0.01 & -- \\
148 & BLHer & 0.42679 & LMC0533-6920C & 52969.042 & 0.572 & 17.13 & 0.02 & 16.85 & 0.03 & 16.66 & 0.07 & $-0.292$ \\
148 & BLHer & 0.42679 & LMC0533-6940I & 52226.973 & 0.824 & 16.84 & 0.02 & 16.67 & 0.04 & 16.57 & 0.10 & $ 0.110$ \\
149 & RVTau & 1.62819 & LMC0533-6940H & 52218.098 & 0.615 & 13.67 & 0.01 & 13.36 & 0.01 & 13.26 & 0.02 & -- \\
150 & WVir  & 0.73981 & LMC0534-7100F & 52563.936 & 0.352 & 15.77 & 0.02 & 15.53 & 0.02 & 15.38 & 0.04 & $-0.021$ \\
151 & WVir  & 0.89693 & LMC0534-7000D & 52314.930 & 0.863 & 15.87 & 0.02 & 15.46 & 0.02 & 15.39 & 0.04 & $ 0.040$ \\
152 & WVir  & 0.96918 & LMC0534-7000D & 52314.930 & 0.938 & 15.55 & 0.02 & 15.15 & 0.02 & 15.05 & 0.03 & $ 0.180$ \\
153 & BLHer & 0.07010 & LMC0534-7100B & 52563.994 & 0.623 & 16.55 & 0.03 & 16.47 & 0.03 & 16.35 & 0.07 & $-0.048$ \\
154 & pWVir & 0.87955 & LMC0534-7100G & 52563.922 & 0.498 & 15.04 & 0.02 & 14.86 & 0.01 & 14.77 & 0.03 & -- \\
155 & WVir  & 0.83871 & LMC0534-7040A & 52335.748 & 0.926 & 15.70 & 0.04 & 15.23 & 0.05 & 15.12 & 0.05 & $ 0.049$ \\
156 & WVir  & 1.18714 & LMC0534-7100D & 52563.965 & 0.886 & 15.03 & 0.02 & 14.67 & 0.01 & 14.55 & 0.02 & $ 0.267$ \\
157 & WVir  & 1.15639 & LMC0537-6920C & 52958.909 & 0.289 & 15.36 & 0.01 & 14.90 & 0.01 & 14.78 & 0.02 & $-0.173$ \\
158 & WVir  & 0.85365 & LMC0536-6820B & 53364.839 & 0.694 & 16.12 & 0.01 & 15.73 & 0.02 & 15.69 & 0.06 & $-0.011$ \\
159 & WVir  & 0.82122 & LMC0537-6940H & 53019.932 & 0.464 & 16.11 & 0.01 & 15.72 & 0.02 & 15.59 & 0.03 & $-0.038$ \\
160 & BLHer & 0.24469 & LMC0538-7100I & 52514.153 & 0.627 & 17.49 & 0.03 & 17.11 & 0.06 & 17.05 & 0.09 & $-0.243$ \\
161 & WVir  & 0.93107 & LMC0539-7120I & 53334.830 & 0.376 & 15.18 & 0.01 & 14.71 & 0.02 & 14.59 & 0.03 & $-0.045$ \\
162 & RVTau & 1.48279 & LMC0537-7000H & 52314.847 & 0.942 & 14.12 & 0.02 & 13.69 & 0.02 & 13.53 & 0.02 & -- \\
163 & BLHer & 0.22881 & LMC0540-6820F & 53355.951 & 0.379 & 17.12 & 0.02 & 16.73 & 0.03 & 16.74 & 0.12 & $ 0.041$ \\
164 & pWVir & 0.92918 & LMC0538-7020E & 52315.872 & 0.174 & 15.25 & 0.02 & 14.82 & 0.02 & 14.50 & 0.02 & -- \\
165 & BLHer & 0.09371 & LMC0537-6920A & 52954.932 & 0.736 & 18.09 & 0.04 & 17.83 & 0.09 &    -- &   -- & $-0.098$ \\
166 & BLHer & 0.32441 & LMC0537-6940A & 53331.928 & 0.098 & 16.23 & 0.02 & 15.88 & 0.02 & 15.79 & 0.03 & $ 0.092$ \\
167 & BLHer & 0.36395 & LMC0541-6940F & 53276.054 & 0.232 & 17.01 & 0.02 & 16.65 & 0.03 & 16.49 & 0.05 & $ 0.068$ \\
168 & WVir  & 1.19585 & LMC0540-6820E & 53355.939 & 0.295 & 15.22 & 0.01 & 14.74 & 0.01 & 14.62 & 0.02 & $-0.199$ \\
169 & RVTau & 1.49074 & LMC0539-7120D & 53334.783 & 0.534 & 14.60 & 0.02 & 14.14 & 0.01 & 14.01 & 0.03 & $-0.451$ \\
170 & WVir  & 0.88553 & LMC0540-6840B & 52691.787 & 0.524 & 16.05 & 0.02 & 15.61 & 0.02 & 15.51 & 0.04 & $-0.071$ \\
171 & BLHer & 0.19166 & LMC0541-7000F & 52306.813 & 0.553 & 17.44 & 0.05 & 17.16 & 0.08 & 17.03 & 0.20 & $-0.141$ \\
172 & WVir  & 1.05002 & LMC0541-7000C & 52305.912 & 0.382 & 15.58 & 0.02 & 15.16 & 0.02 & 15.02 & 0.04 & $-0.138$ \\
173 & WVir  & 0.61783 & LMC0541-7000I & 52306.863 & 0.531 & 17.15 & 0.03 & 16.55 & 0.06 & 16.34 & 0.11 & $-0.100$ \\
174 & RVTau & 1.67042 & LMC0541-6940F & 53276.054 & 0.703 & 13.16 & 0.02 & 12.73 & 0.01 & 12.11 & 0.01 & -- \\
175 & WVir  & 0.96968 & LMC0540-6840H & 52691.886 & 0.215 & 15.74 & 0.02 & 15.31 & 0.02 & 15.18 & 0.03 & $ 0.011$ \\
176 & WVir  & 0.90253 & LMC0539-6800D & 52683.885 & 0.226 & 15.77 & 0.02 & 15.37 & 0.02 & 15.32 & 0.04 & $ 0.007$ \\
177 & WVir  & 1.17713 & LMC0541-6920H & 52528.126 & 0.298 & 15.33 & 0.01 & 14.83 & 0.01 & 14.67 & 0.01 & $-0.167$ \\
178 & WVir  & 1.08680 & LMC0542-7020B & 52306.897 & 0.456 & 15.71 & 0.02 & 15.28 & 0.02 & 15.18 & 0.03 & $-0.144$ \\
179 & WVir  & 0.90580 & LMC0541-7000D & 52306.767 & 0.133 & 15.91 & 0.02 & 15.50 & 0.02 & 15.36 & 0.04 & $ 0.047$ \\
180 & RVTau & 1.49131 & LMC0544-6840H & 52685.857 & 0.394 & 14.03 & 0.02 & 13.57 & 0.01 & 13.08 & 0.02 & -- \\
181 & pWVir & 0.85809 & LMC0542-7040D & 52312.856 & 0.308 & 15.57 & 0.02 & 15.35 & 0.02 & 15.17 & 0.03 & -- \\
182 & WVir  & 0.91521 & LMC0542-7040D & 52312.856 & 0.191 & 15.52 & 0.01 & 15.21 & 0.01 & 15.06 & 0.02 & $ 0.132$ \\
183 & WVir  & 0.81356 & LMC0545-6940B & 52293.036 & 0.909 & 16.39 & 0.02 & 15.88 & 0.03 & 15.80 & 0.05 & $ 0.047$ \\
184 & WVir  & 1.17145 & LMC0545-6940H & 53427.831 & 0.446 & 15.84 & 0.02 & 15.29 & 0.02 & 15.17 & 0.04 & $-0.309$ \\
185 & WVir  & 1.10341 & LMC0545-7000H & 52305.849 & 0.008 & 13.60 & 0.01 & 13.12 & 0.02 & 12.97 & 0.02 & $ 0.066$ \\
186 & WVir  & 1.21383 & LMC0544-6840A & 52684.782 & 0.109 & 14.79 & 0.02 & 14.45 & 0.01 & 14.37 & 0.02 & $ 0.309$ \\
187 & BLHer & 0.38102 & LMC0545-6940A & 53423.853 & 0.068 & 16.98 & 0.03 & 16.73 & 0.07 & 16.72 & 0.15 & $ 0.249$ \\
188 & BLHer & 0.02091 & LMC0545-7000D & 52303.874 & 0.885 & 17.60 & 0.06 & 17.54 & 0.09 &    -- &   -- & $ 0.061$ \\
189 & BLHer & 0.11649 & LMC0547-6840B & 53722.781 & 0.868 & 17.50 & 0.05 & 17.32 & 0.08 & 16.96 & 0.16 & $ 0.140$ \\
190 & RVTau & 1.58390 & LMC0549-6940G & 53491.741 & 0.735 & 13.61 & 0.02 & 13.28 & 0.01 & 13.19 & 0.03 & -- \\
191 & RVTau & 1.53586 & LMC0553-7000I & 52362.918 & 0.173 & 13.60 & 0.02 & 13.27 & 0.01 & 13.05 & 0.02 & -- \\
192 & RVTau & 1.41820 & LMC0554-7020E & 52345.935 & 0.299 & 14.66 & 0.02 & 14.27 & 0.02 & 14.16 & 0.02 & $-0.183$ \\
193 & WVir  & 0.84540 & LMC0557-7200I & 52313.831 & 0.794 & 15.96 & 0.02 & 15.60 & 0.02 & 15.48 & 0.05 & $ 0.030$ \\
\hline
\end{tabular}
\end{center}
\end{minipage}
\end{table*}

\end{document}